\documentclass[a4paper,10pt,twoside]{report}

\usepackage{geometry}	
\usepackage{setspace}

\everydisplay{\def\arraystretch{0.8}}

\usepackage{tocloft}

\renewcommand\thesection{\arabic{section}}

\makeatletter
\def\hlinewd#1{%
	\noalign{\ifnum0=`}\fi\hrule \@height #1 %
	\futurelet\reserved@a\@xhline}
\makeatother

\usepackage{verbatim}
 
\usepackage{hyperref}
\hypersetup{colorlinks=false}
\usepackage{array}

\usepackage[latin9,utf8]{inputenc} 
\usepackage[T1]{fontenc}

\usepackage{cancel}
\usepackage{float,hypcap}
\usepackage{changepage,titlesec}
\usepackage{sectsty}
\usepackage{soul}
\usepackage{bbold}
\usepackage{slashed}
\usepackage{tabularx}
\usepackage{amsmath,amsfonts,amssymb}
\usepackage{amsmath,bbm,latexsym,amssymb}
\usepackage{braket}
\usepackage{blkarray}
\usepackage{enumerate}
\usepackage{booktabs,hyperref}
\usepackage{bigints}
\usepackage{cite}
\usepackage{mflogo}
\usepackage{scrextend}
\usepackage{longtable}
\usepackage{enumitem}
\usepackage{pifont}
\usepackage{cases}
\usepackage{colortbl}
\usepackage[export]{adjustbox}
\usepackage{arydshln}
\usepackage{colortbl}
\usepackage{graphics}

\usepackage{multirow}

\usepackage[flushleft]{threeparttable}

\usepackage{verbatim}

\usepackage{fancyvrb}
\usepackage{xcolor}

\usepackage{graphicx}
\usepackage{caption}
\usepackage{subfig}
\usepackage{feynmp-auto}

\usepackage{color}
\definecolor{orange}{RGB}{255, 222, 173}
\definecolor{uglyblue}{RGB}{95,158,160}
\definecolor{newblue}{RGB}{128,0,0}
\definecolor{mygray}{RGB}{220,220,220}
\definecolor{mywhite}{RGB}{255,250,240}
\usepackage[most]{tcolorbox}

\let\oldcdot\cdot 
\usepackage{breqn} 
\let\cdot\oldcdot 
\usepackage{authblk}

\allowdisplaybreaks

\def\be{\begin{equation}}
\def\ee{\end{equation}}
\def\ba{\begin{alignedat}}
\def\ea{\end{alignedat}}
\def\bea{\begin{eqnarray}}
\def\eea{\end{eqnarray}}
\newcommand{\bs}{\begin{subequations}}
\newcommand{\es}{\end{subequations}}

\def\vs{\vspace}

\def\no{\nonumber\\}
\def\fn{\footnote}

\newcommand{\newc}{\newcommand}
\newc{\ol}{\overline}
\newc{\wt}{\widetilde}

\newc{\m}{\mathcal}

\newcommand{\beq}{\begin{eqnarray}}
\newcommand{\eeq}{\end{eqnarray}}
\newcommand{\bpmatrix}{\begin{pmatrix}}
\newcommand{\epmatrix}{\end{pmatrix}}

\renewcommand{\ol}{\text{1l}}

\renewcommand{\eqref}[1]{eq.~(\ref{#1})}

\newcommand{\bc}{\begin{center}}
\newcommand{\ec}{\end{center}}

\newcommand{\gsim}{\raisebox{-0.13cm}{~\shortstack{$>$ \\[-0.07cm]
      $\sim$}}~}

\def\m#1{m_{#1}}

\def\ltap{\;\centeron{\raise.35ex\hbox{$<$}}{\lower.65ex\hbox{$\sim$}}\;}
\def\gtap{\;\centeron{\raise.35ex\hbox{$>$}}{\lower.65ex\hbox{$\sim$}}\;}

\usepackage{stackengine}
\stackMath

\usepackage{lscape}

\usepackage{imakeidx}
\makeindex
\usepackage{makecell}

\newcommand{\mF}{\mathcal{F}}
\newcommand{\mG}{\mathcal{G}}
\newcommand{\mL}{\mathcal{L}}

\newcommand{\PCR}{\textrm{PC}^{\mathrm{R}}}
\newcommand{\PCC}{\textrm{PC}^{\mathrm{C}}}
\newcommand{\PCT}{\textrm{PC}^{\mathrm{T}}}

\definecolor{neworange}{rgb}{1,0.5,0}

\definecolor{mygreen}{RGB}{0,128,0}

\definecolor{raq}{RGB}{255,164,0}

\newcommand{\cba}{c_{\beta-\alpha}}
\newcommand{\sba}{s_{\beta-\alpha}}
\newcommand{\mbarh}{m_{12}^2}

\begin{document}

\begin{flushright}
IPARCOS-UCM-23-1290    
\end{flushright}

\pagestyle{plain}
\hypersetup{pageanchor=false}
\hypersetup{pageanchor=true}
\pagenumbering{roman}
\setcounter{page}{1}
\pagenumbering{arabic}
\setcounter{page}{1}

\makeatletter         
\renewcommand\maketitle{
{\raggedright 
\begin{center}
{\large \bfseries \@title }

\vspace{1mm}

{\@author}

\vspace{1.2mm}

\end{center}}} 
\makeatother

\title{Is the HEFT matching unique?}
\renewcommand*{\thefootnote}{\fnsymbol{footnote}}
\author[1]{Sally Dawson\fn{dawson@bnl.gov}}
\author[1]{Duarte Fontes\fn{dfontes@bnl.gov}}
\author[2]{Carlos Quezada-Calonge\fn{cquezada@ucm.es}}
\author[2]{Juan Jos\'e Sanz-Cillero\fn{jjsanzcillero@ucm.es}}
\affil[1]{\textit{Department of Physics, Brookhaven National Laboratory, Upton, New York 11973 U.S.A.}}
\affil[2]{\textit{Departamento de Física Teórica and IPARCOS, Universidad Complutense de Madrid, \linebreak Plaza de las Ciencias 1, 28040-Madrid, Spain}}

\date{\today}

\maketitle

\renewcommand*{\thefootnote}{\arabic{footnote}}
\setcounter{footnote}{0}

\vs{0.5mm}
\begin{addmargin}[12mm]{12mm}
\small

Physics beyond the Standard Model (BSM) can be described in a consistent and general way through the Higgs Effective Field Theory (HEFT). Measurements of model-independent HEFT coefficients allow one to constrain the parameter space of BSM models via a matching procedure. In this work, we show that this procedure is not unique and depends on the scalings of the parameters of the Lagrangian. As examples, we consider three BSM models: the real singlet extension of the SM with a $Z_2$ symmetry, the complex singlet extension (CSE) of the SM and the 2 Higgs Doublet Model. We discuss several physical observables, and show that different scalings of the model parameters with the UV scale in the matching to the HEFT can yield quite different results. This complicates the interpretation of HEFT measurements in terms of parameters of BSM models. Additionally, as a by-product, we report the first matching of the CSE to the HEFT.
\end{addmargin}

\normalsize

\section{Introduction}
\label{sec:Introduction}

The discovery of a scalar particle at the Large Hadron Collider (LHC) in 2012 \cite{Aad:2012tfa,Chatrchyan:2012ufa} --- usually identified as the last piece of the Standard Model (SM), the Higgs boson --- intensified an old question: is there physics beyond the SM (BSM)? The absence of any detection of new physics at the LHC suggests that potential BSM physics should be heavy. In that case, the general and consistent framework of an Effective Field Theory (EFT) is ideal to parametrize possible deviations of the SM at the LHC. The two main EFTs for physics that could affect the Higgs sector are the Standard Model EFT (SMEFT)~\cite{Weinberg:1979sa,Buchmuller:1985jz,Leung:1984ni} 
and the Higgs EFT (HEFT)~\cite{Cohen:2020xca,Feruglio:1992wf,Bagger:1993zf,Koulovassilopoulos:1993pw,Buchalla:2013rka,Alonso:2012px,Krause:2018cwe,Brivio:2013pma}; for an introduction to both and to their comparison, see refs.~\cite{Brivio:2017vri,LHCHiggsCrossSectionWorkingGroup:2016ypw}.

Any EFT is constructed on the basis of an expansion; the set of rules that organizes the order of such expansion --- by determining how the different quantities scale --- is known as \textit{power counting} (PC). Between the SMEFT and the HEFT, the former is more widely used given its clear PC  \cite{Brivio:2017vri}.
Indeed, the SMEFT is the default choice in many LHC analyses and allows for the inclusion of data from many different processes (Higgs physics, di-boson production, electroweak precision measurements and top quark physics, among others) \cite{BarrancoNavarro:2021wek,Sciandra:2023qak,Soldatov:2023gwt}. On the other hand, the HEFT is more general than the SMEFT, and has been the object of 
diverse studies in recent years~\cite{
Eboli:2021unw,Criado:2021tec,Cohen:2021ucp,Asiain:2021lch,Alonso:2021rac,Gomez-Ambrosio:2022qsi,Gomez-Ambrosio:2022why,Herrero:2022krh,Domenech:2022uud,Lindner:2022kxm,Quezada-Calonge:2022lop,Graf:2022rco,Sun:2022aag,Sun:2022ssa,Sun:2022snw,Dong:2022jru,Liu:2023jbq,Alonso:2023upf,Asiain:2023myt,Dawson:2023ebe,Arco:2023sac,Bhardwaj:2023ufl,Delgado:2023ynh,Eboli:2023mny,Herrero:2021iqt,Delgado:2017cls},  
extending previous LHC analyses of leading and subleading HEFT couplings~\cite{ATLAS:2014jzl,ATLAS:2016nmw,Espriu:2013fia,Delgado:2013hxa,Delgado:2014jda,Buchalla:2015qju,Fabbrichesi:2015hsa,Brivio:2013pma,Brivio:2016fzo,deBlas:2018tjm}.
The larger freedom provided by the HEFT has also recently motivated the ATLAS collaboration at the LHC to perform an analysis utilizing the HEFT, in the context of double Higgs production \cite{ATLAS:2022fxe}. This suggests that future experimental analyses related to the exploration of Higgs self-interactions may also find  the HEFT parametrization useful.

However, the HEFT is a weak scale effective description of some complete ultraviolet (UV) model. Ultimately, then, one will need to convert the HEFT parametrization to that of a particular UV model. The procedure that allows one to relate the EFT to some higher-energy theory is known as \textit{matching}. Matching is essential both in the scenario where non-zero HEFT coefficients are observed (in which case one would have found BSM physics, and would  seek a UV model that could accommodate it), as well as in the scenario where all HEFT coefficients are consistent with the SM (in which case one uses the contraints of the HEFT coefficients to constrain the parameter space of UV models). The HEFT has often been matched to composite Higgs models \cite{Grojean:2013qca,Alonso:2014wta,Hierro:2015nna,Gavela:2016vte,Qi:2019ocx,Lindner:2022kxm}. Up to our knowledge, in 2016 ref. \cite{Buchalla:2016bse} performed for the first time the matching of the HEFT to a UV complete model, by considering a $Z_2$-symmetric real singlet extension (Z2RSE) of the SM. Very recently, the same exercise has been performed in the 2 Higgs Doublet Model (2HDM) \cite{Dawson:2023ebe,Arco:2023sac}.  

Naively, the matching between the HEFT and a UV model with heavy particles is done simply by performing an expansion in inverse powers of the heavy physical masses. This seems to define an unambiguous expansion, i.e. a unique PC.
In this paper, we question this reasoning. We show that the existence of a multiplicity of choices for independent parameters complicates the problem. In fact, even if one adopts an expansion simply in inverse powers of the heavy physical masses, very different PCs are obtained according to the set of parameters taken as independent, as different sets imply different scalings. More interestingly, instead of happening that a single PC is always most adequate for all observables, one should in general use different PCs to ensure a fast replication of the results in the UV model for different processes or different regions of the parameter space.
This complicates the interpretation of HEFT measurements in terms of parameters 
of UV models.%
\fn{For details on the low-energy HEFT expansion, see refs.~\cite{Weinberg:1978kz,Manohar:1983md,Jenkins:2013sda,Buchalla:2013eza,Pich:2016lew,Gavela:2016bzc}.}
We illustrate these features by considering three UV models: the Z2RSE, the complex singlet extension (CSE) of the SM and the 2HDM. For each one, we present three PCs, and study the quality with which they replicate the full model in different observables.

The paper is organized as follows: we start by presenting the models and the PCs in Section~\ref{sec:models}. In this section, we also discuss how to compare the SMEFT and the HEFT matchings, and we write all matching relations. We then present our results in Section~\ref{sec:results}, and we summarize our conclusions in Section~\ref{sec:conclusions}. The Appendix provides extra details about the the HEFT approach to the CSE.

\section{Models, countings and matchings}
\label{sec:models}

In this section, we describe the models discussed in this article. For each case, we provide the Lagrangian, discuss relevant choices of independent parameters and relevant Feynman rules, describe the restrictions to the parameter space of the model, and present the PCs to be investigated. Some introductory notes are in order: 
\begin{enumerate}
\item In all the three models considered in this paper, we assume that all the particles beyond the SM are heavy, so that the SM can represent an EFT of the model at stake.
\item The notion of PC is usually employed in the literature from a bottom-up perspective, where the EFT is not bound to any particular UV completion (for an excellent discussion, see ref. \cite{Brivio:2017vri}). Here we use the term in a broader sense, to describe any set of rules that fixes the different orders of the EFT expansion. In particular, 
the notion of PC in that sense characterizes a certain matching between a UV model and the HEFT. Different matchings between the UV model and the HEFT therefore correspond to different PCs.

\item We follow ref. \cite{Dawson:2023ebe} in organizing the PCs via a small auxiliary parameter $\xi$, corresponding to an inverse heavy scale (see also ref. \cite{Dittmaier:2021fls}).
More specifically, if $v \sim 250$ GeV represents the electroweak scale and $\Lambda \sim 800$ GeV some heavy scale, we estimate $\xi \sim (v/\Lambda)^2 \sim 0.1$.
The effective Lagrangian, being an expansion in inverse powers of the heavy scale, can thus be written as an expansion in non-negative powers of $\xi$.
All PCs discussed in this paper lead to well-defined expansions, in the sense that the effective Lagrangians do not contain negative powers of $\xi$ (i.e. positive powers of the large scale).
In a certain model, the different PCs differ in the way they attribute different scalings in powers of $\xi$ to the parameters. In general, the different scalings do not have a specific physical motivation; they are considered solely with the purpose of illustrating different possibilities of matching.

\item It should be clear that we are not performing an exhaustive study; that is, we will not consider all possible PCs that could be considered. Rather, the goal is to illustrate the multiplicity of possible matchings between HEFT and a given UV model, and to discuss relevant features of that multiplicity.

\item We shall present in all models the \textit{decoupling} PC. This is the PC consistent with the decoupling limit of the model at stake. In other words, the decoupling PC is such that, on the one hand, the trivial order ($\xi^0$) of the effective Lagrangian is the SM Lagrangian and, on the other, the quartic parameters of the potential do not scale with the heavy masses, in accordance with perturbative unitarity (for a recent discussion, see ref. \cite{Dawson:2023ebe}).%
\fn{In principle, there can be more than one decoupling PC for each model. The investigation of this possibility is beyond the scope of this paper.}
Then, for the decoupling PC, the $\mathcal{O}(\xi^1)$ and $\mathcal{O}(\xi^2)$ terms of the effective Lagrangian  correspond to the SMEFT dimension-6 and dimension-8 contributions in that model, respectively.
Such a coincidence of results has been already observed in the case of the 2HDM in ref. \cite{Dawson:2023ebe}.
For simplicity, the decoupling PC of each model will always be represented by the same color (yellow) in all the plots of this paper.

\item For each PC in each model, we provide the matching of the relevant parameters of the HEFT Lagrangian up to $\mathcal{O}(\xi^2)$.

\end{enumerate}

\noindent We now turn to the HEFT
Lagrangian, which contains an expansion in the number of (covariant) derivatives. At the leading order (LO) in that expansion, the terms of the Lagrangian which will matter for the processes discussed in this paper are
\be
\mL_{\rm HEFT}
\supset
\dfrac{v^2}{4} \mF(h) {\rm Tr}\left\{D_\mu U^\dagger D_\mu U\right\} 
+ \dfrac{1}{2}(\partial_\mu h)^2 - V(h)   
,
\label{eq:heftdef}
\ee  
where $v=246$ GeV is the vacuum expectation value (vev) of the SM Higgs field, $D_\mu$ is the covariant derivative, and $\mF(h)$ and $V(h)$ are analytical functions, each one consisting of a series in powers of $h$.
In general, one has $D_\mu U= \partial_\mu U + i g W_\mu^a\dfrac{\sigma^a}{2}U-i g' U \dfrac{\sigma^3}{2}B_\mu $, with $U=1$ in the particular case of the unitary gauge. In the HEFT, $h$ is a gauge singlet, which allows $\mF(h)$ and $V(h)$ to contain arbitrary powers of $h$:
\be
\mathcal{F}(h) = 1 + 2 a \dfrac{h}{v} + b \dfrac{h^2}{v^2} + ...\, ,
\qquad
V(h) = \dfrac{1}{2} m_h^2 h^2 \left( 1 +\kappa_3 \dfrac{h}{v} +\dfrac{\kappa_4}{4}\dfrac{h^2}{v^2} \, +... \right)\, , 
\label{eq:flare}
\ee
where $m_h$ is the $h$ mass, $a$, $b$, $\kappa_3$ and $\kappa_4$ are arbitrary HEFT couplings, and the dots stand for terms with higher powers of $h$. These couplings are normalized in such a way that the SM limit is obtained for $a=b=\kappa_3=\kappa_4=1$.

In the case of the 2HDM, we shall also consider the processes $h \to b \bar{b}$ at tree-level, as well as $h \to \gamma \gamma$ and $h \to \gamma Z$ at one-loop (where the existence of a charged Higgs boson has interesting consequences for the EFT). For $h \to b \bar{b}$, we need to consider the term:
\be
\mL_{\rm HEFT}
\supset
- \mathcal{G}(h) \, m_b \, \bar{b} b,
\ee  
where $m_b$ is the $b$ quark mass and $\mG(h)$ is another polynomial function of $h$, such that:
\be
\mG(h) = 1 + c_1 \dfrac{h}{v} + ... \, ,
\label{eq:flare2}
\ee
with $c_1$ another HEFT coupling, with $c_1=1$ in the SM. 
As for $h \to \gamma \gamma$ and $h \to \gamma Z$, in order for these processes to be matched to the HEFT, it is necessary to consider the next to leading order (NLO) terms in the derivative expansion of $\mL_{\rm HEFT}$. We follow ref. \cite{Arco:2023sac} to write those terms as:
\be
\mL_{\rm HEFT} \supset - a_{H B B} \frac{h}{v}
\operatorname{Tr}\left[\hat{B}_{\mu \nu} \hat{B}^{\mu \nu}\right]
- a_{H W W} \frac{h}{v} \operatorname{Tr}\left[\hat{W}_{\mu \nu} \hat{W}^{\mu \nu}\right] + a_{H 1} \frac{h}{v} \operatorname{Tr}\left[U \hat{B}_{\mu \nu} U^{\dagger} \hat{W}^{\mu \nu}\right],
\ee
where $a_{H B B}$, $a_{H W W}$ and $a_{H 1}$ are yet other HEFT couplings (which vanish in the SM limit), and:
\be
\hat{B}_\mu =g^{\prime} B_\mu \dfrac{\sigma^3}{2},
\quad
\hat{W}_\mu = g W_\mu^a \dfrac{\sigma^a}{2},
\quad
\hat{B}_{\mu \nu} = \partial_\mu \hat{B}_\nu-\partial_\nu \hat{B}_\mu,
\quad
\hat{W}_{\mu \nu}=\partial_\mu \hat{W}_\nu-\partial_\nu \hat{W}_\mu+i\left[\hat{W}_\mu, \hat{W}_\nu\right].
\ee
Then, the relevant couplings for $h\to \gamma\gamma$ and $h \to \gamma Z$ are, respectively,
\be
a_{h \gamma \gamma} =a_{H B B}+a_{H W W}-a_{H 1},
\qquad
a_{h \gamma Z} = a_{H B B} + a_{H W W} - a_{H 1} + \dfrac{m_Z^2}{2 m_W^2} \left(a_{H 1} - 2 a_{H B B}\right).
\ee
Once the matching to HEFT has been performed, we will express some matching results as $\Delta x= x - 1$, where $x$ represents a generic HEFT coupling. The SM limit is recovered when $\Delta x=0$. The couplings that we consider in this work are collected in tables~\ref{tab:coeffs_Z2RSE}, \ref{tab:coeffs_CSE}, \ref{tab:coeffs_2HDM_a} and~\ref{tab:coeffs_2HDM_b}.

Finally, a note on the relation between the HEFT and the SMEFT. Since the HEFT is more general than SMEFT, it is always possible to write any SMEFT in HEFT form, but the opposite path is not always guaranteed. It is thus interesting to determine when a given HEFT can be written in SMEFT coordinates. Recently, new approaches have emerged where EFTs are formulated from a geometrical point of view~\cite{Alonso:2015fsp,Alonso:2016oah,Assi:2023zid,Trott:2017yhn,Cohen:2020xca}. This framework allows to ascertain whether a certain HEFT can be written in SMEFT coordinates by studying the curvature of the EFT manifold. The gist of the idea is to study the existence of an O(4) fixed point upon which the SMEFT can be formulated. Since the purpose of this paper is to illustrate different matchings in HEFT for multiple UV models, we will not further explore this geometrical interpretation.

\subsection{Z2RSE}
\label{sec:Z2RSE-model}

\subsubsection{The model}

We start with the simplest extension, Z2RSE. 
We present here a short review of the model, following ref. \cite{Buchalla:2016bse} (for details, cf. e.g. ref. \cite{Robens:2015gla,Robens:2016xkb}).  
The model is obtained by taking the SM (whose Higgs doublet we identify as $\phi$) and adding a real singlet $S$ to it, which transforms as $S \to -S$ under $Z_2$. The potential reads:
\be
\label{eq:Z2RSE-V}
V=-\dfrac{\mu_1^2}{2} \phi^{\dagger} \phi-\dfrac{\mu_2^2}{2} S^2+\dfrac{\lambda_1}{4}\left(\phi^{\dagger} \phi\right)^2+\dfrac{\lambda_2}{4} S^4+\dfrac{\lambda_3}{2} \phi^{\dagger} \phi S^2,
\ee
where all the parameters are real. The fields $\phi$ and $S$ may be parametrized as:
\be
\phi=\left(\begin{array}{c}
G^{+} \\
\dfrac{1}{\sqrt{2}}\left(v+h_1 + i G_0\right)
\end{array}\right),
\quad
S= \dfrac{v_s + h_2}{\sqrt{2}},
\ee
where $G^+$ and $G_0$ are respectively the charged and neutral would-be Goldstone bosons, $h_1$ and $h_2$ are neutral scalar fields (not yet mass states), and $v=246 \, \mathrm{GeV}$ and $v_s$ are vevs, which we assume to be real without loss of generality. 
The minimization equations read:
\be
\label{eq:Z2RSE_minima}
\mu_1^2=\dfrac{\lambda_1 v^2+\lambda_3 v_s^2}{2},
\qquad
\mu_2^2=\dfrac{\lambda_3 v^2+\lambda_2 v_s^2}{2}.
\ee
The mass states $h$ and $H$ --- with masses $m=125 \, \mathrm{GeV}$ and $M$, respectively --- are obtained from $h_1$ and $h_2$ by considering a mixing angle $\chi$, such that:
\be
\left(\begin{array}{c}
h \\
H
\end{array}\right)
=
\left(\begin{array}{cc}
c_\chi & - s_\chi \\
s_\chi & c_\chi
\end{array}\right)\left(\begin{array}{c}
h_1 \\
h_2
\end{array}\right),
\ee
where we introduced the  notation $s_x \equiv \sin x, c_x \equiv \cos x$.
The model thus contains, besides the SM particles, a heavy real scalar $H$. One possible set of independent parameters is:
\be
\label{eq:1-indep-set-Z2RSE}
v, \, m, \, v_s, \, M, \, s_{\chi}.
\ee
The relations between $\lambda_1$, $\lambda_2$ and $\lambda_3$ of eq.~(\ref{eq:Z2RSE-V}) and the parameters of eq.~(\ref{eq:1-indep-set-Z2RSE}) read:
\bs
\label{eq:Z2RSE-lambdas}
\bea
\label{eq:Z2RSE-lambdas-1}
\lambda_1 &=& \frac{2}{v^2}\left[M^2 s_\chi^2-m^2\left(s_\chi^2-1\right)\right] \\
\label{eq:Z2RSE-lambdas-2}
\lambda_2 &=& \frac{2}{v_s^2}\left[m^2 s_\chi^2-M^2\left(s_\chi^2-1\right)\right] \\
\label{eq:Z2RSE-lambdas-3}
\lambda_3 &=& \frac{2 c_\chi s_\chi}{v \, v_s}\left(M^2-m^2\right).
\eea
\es
Another relevant choice uses eq.~(\ref{eq:Z2RSE_minima}) to replace $v_s$ by $\mu_2^2$:
\be
\label{eq:2-indep-set-Z2RSE}
v, \, m, \, \mu_2^2, \, M, \, s_{\chi}.
\ee

\subsubsection{Restrictions}

The parameter space of the Z2RSE is restricted by many constraints, both purely theoretical and experimental.  Theoretical consistency requires that the scattering amplitudes satisfy perturbative unitarity, that the couplings in the Lagrangian are perturbative, and that the parameters correspond to a minimum of the potential\cite{Dawson:2021jcl,
Ilnicka_2018}.  On the experimental side, the  limits come from precision electroweak measurements, including the W boson mass\cite{L_pez_Val_2014,Dawson:2021jcl}, measurements of the Higgs coupling strengths\cite{Englert:2014uua}, and direct searches for the heavy Higgs boson of the model.   Higgs coupling measurements require $|s_{\chi}| \lesssim 0.3$ for $m \lesssim 2 M$.  For $M \lesssim 850$ GeV, the most stringent limit comes from the measurement of the W boson mass, while for $M\gsim 850$ GeV the strongest limit is from the requirement that the quartic couplings remain perturbative ($\lambda_{1,2,3} < 8 \pi$)\cite{Robens:2022oue}.  The limit from the W boson mass is independent of $v_s$.
Direct searches for the heavy Higgs boson in single $H$ production, as well as searches for the resonant process $pp\rightarrow H\rightarrow hh$, do not significantly change the bounds for $M$ between $600-800$ GeV \cite{Robens:2022oue}, which are the the typical scales chosen in our plots. We will only consider $|s_{\chi}| \lesssim 0.2$, which is allowed by all the limits listed here.

\subsubsection{Power countings}

We consider the following EFT PCs for the Z2RSE:
\begin{itemize}
\item $\PCR_1$ takes eq.~(\ref{eq:1-indep-set-Z2RSE}) as the set of independent parameters, and imposes the decoupling scaling:
\be
\label{eq:PCR_1}
M^2 \sim \mathcal{O}(\xi^{-1}),
\quad
v_s^2 \sim \mathcal{O}(\xi^{-1}),
\quad
s_{\chi}^2 \sim \mathcal{O}(\xi)\, . 
\ee
\item $\PCR_2$ takes eq.~(\ref{eq:1-indep-set-Z2RSE}) as the set of independent parameters, and imposes:
\be
\label{eq:PCR_2}
M^2 \sim \mathcal{O}(\xi^{-1})\, .  
\ee
\item $\PCR_3$ takes eq.~(\ref{eq:2-indep-set-Z2RSE}) as the set of independent parameters, and imposes:
\be
\label{eq:PCR_3}
M^2 \sim \mathcal{O}(\xi^{-1})\, .
\ee
\end{itemize}
Here and it what follows, any parameter of the chosen set that is not explicitly mentioned is assumed to scale as $\mathcal{O}(\xi^0)$ in that PC. As mentioned above, it is reasonable to use all the three PCs of eqs. \ref{eq:PCR_1} to \ref{eq:PCR_3} in the matching, as they do not lead to positive powers of $M$ in the effective Lagrangian.
$\PCR_1$ was put forward at the end of ref. \cite{Dawson:2023ebe} and is the decoupling PC. In fact, and as discussed in ref. \cite{Dawson:2023ebe}, eqs.~(\ref{eq:Z2RSE-lambdas-1})--(\ref{eq:Z2RSE-lambdas-3})     
show that the quartic parameters in the UV model do not scale as heavy parameters (i.e. with negative powers of $\xi$) if the scaling~(\ref{eq:PCR_1}) is obeyed. We stress that both the scaling of $v_s$ as a heavy parameter and the scaling of $s_{\chi}$ as a small quantity are crucial to this end. In contrast, by restricting themselves to scaling only $M$, both $\PCR_2$ and $\PCR_3$ do lead to negative powers of $\xi$ in the scaling of the quartic parameters. Note that $\PCR_2$ is the PC used in ref. \cite{Buchalla:2016bse} in the context of the HEFT. $\PCR_3$ is identical to $\PCR_2$, except that it takes $\mu_2^2$ instead of $v_s$ as an independent $\mathcal{O}(\xi^0)$ parameter. We will show that this change has significant consequences.
Finally, we checked that, in $\PCR_1$, and for the processes discussed here, the existence of odd powers of $v_s$ or $s_{\chi}$ does not introduce non-integers powers of $\xi$ in the expansion, as odd powers of $v_s$ always multiply odd powers of $s_{\chi}$.

The matching results for the different PCs are shown in table~\ref{tab:coeffs_Z2RSE} up to $\mathcal{O}(\xi^2)$, such that the powers of $\xi$ are explicitly included. It is straightforward to see that all $\Delta x$ deviations vanish in the alignment limit, $s_{\chi}=0$, where the SM couplings are recovered. At LO in $\xi$, $\PCR_2$ yields the results given in ref.~\cite{Buchalla:2016bse}. 
\begin{table}[!h]
\begin{normalsize}
\normalsize
\begin{center}
\begin{tabular}
{>{\centering\arraybackslash}m{0.6cm}
>{\centering\arraybackslash}m{1.5cm}
>{\centering\arraybackslash}m{2.9cm}
>{\centering\arraybackslash}m{3.3cm}
>{\centering\arraybackslash}m{5.5cm}}
\hlinewd{1.1pt}
PC
&
$\Delta a$
&
$\Delta b$
&
$\Delta \kappa_3$
&
$\Delta \kappa_4$\\
\hline\\[-1.5mm]
$\PCR_1$
&
$- \xi \dfrac{s_{\chi}^2}{2} \linebreak -\xi^2 \dfrac{s_{\chi}^4}{8}$
&
$-2 \xi s_{\chi}^2 \linebreak + \xi^2 s_{\chi }^2 \Big(s_{\chi }^2-\dfrac{2 m^2}{M^2} \linebreak - \dfrac{v_s s_{\chi }}{v_s}\Big)$
&
$- \xi \dfrac{3 s_{\chi}^2}{2} +\linebreak  \xi^2 \dfrac{ s_{\chi}^3}{8 v_s} \Big(3 s_{\chi} v_s - 8 v \Big)  $
&
$-\xi \dfrac{25 s_{\chi}^2}{3} - \xi^2 \dfrac{s_{\chi}^2}{3 M^2 v_s} \Big[28 m^2 v_s \linebreak -M^2 s_{\chi} \left(41 s_{\chi} v_s-38 v\right) \Big] $\\ \\\arrayrulecolor{mygray}\hline\\[-1.5mm]
$\PCR_2$
&
$ c_{\chi}-1$
&
$c_{\chi}^4-s_{\chi}^3 c_{\chi} \dfrac{v}{v_s} -1 \linebreak + \xi \dfrac{2 m^2   s_{\chi}^2}{M^2 v_s} \Big(s_{\chi}^2 v_s \linebreak - v_s - c_{\chi} s_{\chi} v  \Big) $
&
$c_{\chi}^3-\dfrac{s_{\chi}^3 v}{v_s}-1$
&
$-1 - \dfrac{19 c_{\chi}^2 s_{\chi}^2 \left(c_{\chi} v_s+s_{\chi} v\right)^2}{3 v_s^2}
\linebreak
+ \dfrac{\left(c_{\chi}^4 v_s^2+s_{\chi}^4 v^2\right)}{v_s^2}
\linebreak 
- \xi \dfrac{28 c_{\chi}^2 m^2 s_{\chi}^2 \left(c_{\chi} v_s+s_{\chi} v\right){}^2}{3 M^2 v_s^2}
\linebreak
-\xi^2 \dfrac{16 c^2 m^4 s_{\chi}^2 \left(c_{\chi} v_s + s_{\chi} v\right){}^2}{3 M^4 v_s^2}$  \\ \\\arrayrulecolor{mygray}\hline\\[-1.5mm]
$\PCR_3$
&
$c_{\chi}-1$
&
$-s_{\chi}^2 + \xi \dfrac{s_{\chi }^2}{M^2} \Big(m^2 \linebreak -\mu _2^2\Big) + \xi^2 \dfrac{3 m^2 s_{\chi }^2}{M^4} \linebreak \times (m^2-\mu_2^2)$
&
$-1+ c_{\chi} \linebreak - 
\xi \dfrac{s_{\chi }^2}{M^2 c_{\chi }} \left(m^2-\mu _2^2\right) \linebreak - \xi^2 \dfrac{m^2s_{\chi }^2}{M^4 c_{\chi}} \left(m^2-\mu _2^2\right)$
&
$-s_{\chi }^2 + 
\xi \dfrac{2 s_{\chi }^2}{M^2} \Big(m^2-\mu _2^2\Big) \linebreak + 
\xi^2 \dfrac{s_{\chi }^2}{3 c_{\chi} M^4} \left(m^2-\mu_2^2\right) \bigg[m^2 \Big(13 s_{\chi }^2\linebreak -10\Big) + \mu_2^2 \left(16-19 s_{\chi }^2\right)\bigg] $\\
\\[-1.5mm]
\hlinewd{1.1pt}
\end{tabular}
\end{center}
\vspace{-5mm}
\end{normalsize}
\caption{HEFT couplings for the Z2RSE. All the couplings are shown up to $\mathcal{O}(\xi^2)$.}
\label{tab:coeffs_Z2RSE}
\end{table}
\normalsize

\subsection{CSE}

\subsubsection{The model}

We now turn to the CSE \cite{Barger:2008jx,Costa:2015llh,Muhlleitner:2017dkd}. We follow ref. \cite{Dawson:2017jja}, and thus include neither a $Z_2$, nor a $U(1)$ symmetry. The starting point is again the SM, and we add to it a complex scalar singlet, $S_c$. We write the potential as:
\bea
&V = -\dfrac{\mu^2}{2} \phi^{\dagger} \phi+\dfrac{\lambda}{4}\left(\phi^{\dagger} \phi\right)^2 + \dfrac{1}{2} b_2 \left|S_c\right|^2 + \dfrac{\delta_2}{2} \phi^{\dagger} \phi\left|S_c\right|^2 + \dfrac{1}{4} d_2 \left(\left|S_c\right|^2\right)^2 + \bigg[a_1 S_c + \dfrac{1}{4} b_1 S_c^2 + \dfrac{1}{6} e_1 S_c^3 & \nonumber \\
& + \dfrac{1}{6} e_2 S_c\left|S_c\right|^2 + \dfrac{1}{8} d_1 S_c^4+\dfrac{1}{8} d_3 S_c^2\left|S_c\right|^2 + \dfrac{1}{4} \delta_1 \phi^{\dagger} \phi S_c + \dfrac{1}{4} \delta_3 \phi^{\dagger} \phi S_c^2 + \textrm{h.c.} \bigg], &
\label{eq:CSE:potential}
\eea
where $\mu^2$, $\lambda$, $d_2$, $\delta_2$ and $b_2$ are real, while the other parameters are complex.
The fields are written as:%
\fn{$S_c$ could have a (complex) vev, but this can be set to zero without loss of generality \cite{Dawson:2017jja}.}
\be
\phi=\left(\begin{array}{c}
G^{+} \\
\dfrac{1}{\sqrt{2}}\left(v + h + i G_0\right)
\end{array}\right),
\quad
S_c= \dfrac{S + i A}{\sqrt{2}},
\ee
where $v$, $G_0$ and $G^{+}$ have the same meaning as for the Z2RSC, and $h$, $S$ and $A$ are real neutral fields, not yet mass states. Requiring them to be free from tadpoles leads to the minimization equations:
\be
\mu^2 = \dfrac{\lambda}{2} v^2,
\qquad
a_1 = - \dfrac{\delta_1}{8} v^2.
\ee
Then, $h$, $S$ and $A$ can be diagonalized introducing the mixing angles $\theta_1$ and $\theta_2$, obeying:%
\fn{The diagonalization matrix could have a third mixing angle, which can be removed without loss of generality \cite{Dawson:2017jja}.}
\be
\left(\begin{array}{l}
h_1 \\
h_2 \\
h_3
\end{array}\right)
=
\left(\begin{array}{ccc}
c_{\theta_1} & -s_{\theta_1} & 0 \\
s_{\theta_1} c_{\theta_2} & c_{\theta_1} c_{\theta_2} & s_{\theta_2} \\
s_{\theta_1} s_{\theta_2} & c_{\theta_1} s_{\theta_2} & -c_{\theta_2}
\end{array}\right)
\left(\begin{array}{l}
h \\
S \\
A
\end{array}\right)
\label{eq:CSE:rotationmatrix},
\ee
where the fields $h_i$ ($i=1,2,3$) are mass states with mass $m_i$, such that $m_1 \equiv m = 125$ GeV. 
%
The model thus contains, besides the SM particles, two heavy real scalars $h_2$ and $h_3$. Taking $\lambda, \delta_1, b_1$ and $b_2$ as dependent parameters, their expressions in terms of the independent parameters read:
\bs
\label{eq:CSE-deprels}
\bea
\label{eq:CSE-lambda}
{\lambda} &=& \dfrac{2}{v^2} \,  \Big[ m_1^2 \, c_1^2 + s_1^2 \,  \left( m_2^2 \, c_2^2 + m_3^2 \, s_2^2 \right) \Big] , \\
{\delta_{1\mathrm{R}}} &=& \dfrac{\sqrt{2} \, s_1 c_1}{v} \Big[m_2^2 + m_3^2 - 2 \, m_1^2 + \left( m_2^2 - m_3^2 \right) (c_2^2 - s_2^2) \Big], \\
{\delta_{1\mathrm{I}}} &=& \dfrac{2 \sqrt{2} \, s_1 s_2 c_2}{v} \left(m_3^2 - m_2^2 \right), \\
{b_{1\mathrm{R}}} &=& - \dfrac{\delta_{3\mathrm{R}} \, v^2}{2} + m_1^2 \, s_1^2 + c_2^2 \left(m_2^2 \, c_1^2 - m_3^2\right) + s_2^2 \left(m_3^2 \, c_1^2 - m_2^2\right), \\
{b_{1\mathrm{I}}} &=& 2 \, c_1 s_2 c_2 \left(m_3^2 -m_2^2\right) - \dfrac{\delta_{3\mathrm{I}} \, v^2}{2}  , \\
{b_2} &=& m_1^2 \, s_1^2 + c_2^2 \left( m_2^2 \, c_1^2 + m_3^2\right) + s_2^2 \left( m_3^2 \, c_1^2 + m_2^2\right) - \dfrac{\delta_{2} \, v^2}{2} ,
\eea
\es
where we used the notation $x_{\mathrm{R}} = \mathrm{Re}(x)$ and $x_{\mathrm{I}} = \mathrm{Im}(x)$, for any $x$. In the following, we take $h_2$ and $h_3$ to be degenerate, and define $M = m_2 = m_3$. The set of independent parameters of the scalar sector then is:
\be
\label{eq:CSE-indep}
v,
\,
m,
M, \theta_1, \theta_2, \delta_2, \delta_3, d_1, d_2, d_3, e_1, e_2.
\ee

\subsubsection{Restrictions}

Similarly to the Z2RSE, we restrict the parameter space of the CSE by taking into account electroweak precision measurements, perturbative unitarity, the perturbativity of the quartic couplings, boundedness of the potential from below, Higgs coupling measurements, and searches for heavy scalars\cite{Egle_2022}. We follow ref. \cite{Adhikari:2022yaa} in using the constraint $|s_1| \lesssim 0.2$ and consider $M \sim 800$ GeV. The parameter points chosen for our numerical results satisfy all of these constraints. We do not make the assumption that one of the scalars is a dark matter candidate as is frequently done\cite{Egle:2023pbm}.

\subsubsection{Power countings}

All the PCs that  we consider for the CSE take eq.~(\ref{eq:CSE-indep}) as the set of independent parameters, such that:
\begin{itemize}
\item $\PCC_1$ imposes the decoupling scaling:
\be
\label{eq:PCC_1}
M^2 \sim \mathcal{O}(\xi^{-1}),
\quad
s_{1} \sim \mathcal{O}(\xi).
\ee
\item $\PCC_2$ imposes:
\be
\label{eq:PCC_2}
M^2 \sim \mathcal{O}(\xi^{-1}),
\quad
s_1^2 \sim \mathcal{O}(\xi),
\quad
e_1^2 \sim \mathcal{O}(\xi^{-1}),
\quad
e_2^2 \sim \mathcal{O}(\xi^{-1}).
\ee
\item $\PCC_3$ imposes:
\be
\label{eq:PCC_3}
M^2 \sim \mathcal{O}(\xi^{-1}).
\ee
\end{itemize}
As before, all these PCs represent consistent approaches to an EFT. The cubic terms, $e_1, e_2$ and $\delta_1$, all have the dimensions of mass in this case. As a consequence, they can potentially scale in different ways in the limit of large $M$. This ambiguity is intrinsically present in the matching of the model to the HEFT. Concerning the three PCs chosen in eqs.~(\ref{eq:PCC_1}) to~(\ref{eq:PCC_3}), $\PCC_1$ is the decoupling PC. From eq.~(\ref{eq:CSE-lambda}), indeed, the scaling~(\ref{eq:PCC_1}) implies that $\lambda$ does not scale with negative powers of $\xi$. This does not happen in $\PCC_2$ or $\PCC_3$. The latter performs an expansion only in inverse powers of $M$. The former not only imposes a stronger scaling for $s_1$ than $\PCC_1$, but also exploits the fact that $e_1$ and $e_2$ are dimensionful parameters to scale them as heavy. We checked that, in the processes discussed here, the trivial order ($\xi^0$) of $\PCC_2$ corresponds to the SM, and the existence of odd powers of $s_{\chi}$, $e_1$ and $e_2$ does not introduce non-integers powers of $\xi$ in the expansion (as odd powers of $e_i$ always multiply odd powers of $s_1$).

The matching results for the different PCs are shown in table~\ref{tab:coeffs_CSE}, where we define  $\bar{\delta}_{23} \equiv \delta_2+\delta_{3\rm{R}}$ and $ e_{12\rm{R}} \equiv e_{1\rm{R}}+e_{2\rm{R}}$. As before, we present the results up to $\mathcal{O}(\xi^2)$ (writing explicitly the powers of $\xi$), except in the cases for which the expressions are too lengthy. In those cases, we indicate the order of the terms not included. The complete matching results up to $\mathcal{O}(\xi^2)$ can be found in the auxiliary file accompanying this manuscript. As in the Z2RSE, all $\Delta x$ vanish in the alignment limit, $s_1 = 0$.
\begin{table}[!h]
\begin{normalsize}
\normalsize
\begin{center}
\begin{tabular}
{
> {\centering\arraybackslash}m{0.6cm}
>{\centering\arraybackslash}m{1.4cm}
>{\centering\arraybackslash}p{1.2cm}
>{\centering\arraybackslash}m{5.0cm}
>{\centering\arraybackslash}m{5.0cm}
}
\hlinewd{1.1pt}
PC
&
$\Delta a$
&
$\Delta b$
&
$\Delta \kappa_3$
&
$\Delta \kappa_4$\\
\hline\\[-1.5mm]
$\PCC_1$
&
$-\xi^2 \dfrac{s_1^2}{2}$
&
$-\xi^2 \hspace{.5mm} 2 s_1^2$
&
$\xi^2 \dfrac{s_1^2}{2 m_1^2} \Big(v^2 \bar{\delta }_{23}-3 m_1^2\Big)$
&
$\xi^2 \dfrac{s_1^2}{3m_1^2} \Big(9 v^2 \bar{\delta }_{23}-25 m_1^2\Big) $ \\ \\\arrayrulecolor{mygray}\hline\\[-1.5mm]
$\PCC_2$
&
$-\xi \dfrac{s_1^2}{2} \linebreak - \xi^2 \dfrac{s_1^4}{8}$
&
$-\xi \, 2 s_1^2 \linebreak + \mathcal{O}(\xi^2)$
&
$-\xi \dfrac{s_1^2}{6 m_1^2} \Big(9 m_1^2 - 3 v^2 \bar{\delta }_{23} \linebreak + \sqrt{2} s_1 v \, e_{12 \rm{R}}\Big) \linebreak +\xi^2 \dfrac{s_1^4}{8 m_1^2} \Big(3 m_1^2-2 v^2 \bar{\delta }_{23}\Big)  $
&
$\xi \dfrac{s_1^2}{3 m_1^2} \Big(9 v^2 \bar{\delta }_{23}-3 \sqrt{2} s_1 v \, e_{12 \rm{R}} \linebreak -25 m_1^2\Big) + \mathcal{O}(\xi^2)$ \\ \\\arrayrulecolor{mygray}\hline\\[-1.5mm]
$\PCC_3$
&
$c_1-1$
&
$c_1^4-1 \linebreak + \mathcal{O}(\xi)$
&
$-\dfrac{v}{24 m_1^2} \bigg[3 c_1 v \big(c_1^2-3 s_1^2-1 \big) \bar{\delta }_{23} \linebreak + \sqrt{2} s_1 \big(-3 c_1^2+s_1^2+3 \big) e_{12 \rm{R}}\bigg] \linebreak + \dfrac{c_1}{4} \Big(c_1^2 -3 s_1^2 +3\Big)-1$
&
$\dfrac{s_1^2 v}{2 m_1^2} \bigg[s_1^2 v \Big(-6 (c_1^2+1) \bar{\delta }_{23}+d_{13 \rm{R}} \linebreak +d_2 \Big)+6 v \bar{\delta }_{23} - 2 \sqrt{2} c_1^3 s_1 e_{12 \rm{R}}\bigg] \linebreak + c_1^4 \bigg( 1-\dfrac{19 s_1^2}{3} \bigg) -1+ \mathcal{O}(\xi) $ \\
\\
\hlinewd{1.1pt}
\end{tabular}
\end{center}
\vspace{-5mm}
\end{normalsize}
\caption{HEFT couplings for the CSE. All the couplings are shown up to $\mathcal{O}(\xi^2)$, except the ones that are too lengthy. The full expressions can be found in the auxiliary file.}
\label{tab:coeffs_CSE}
\end{table}
\normalsize

\subsection{2HDM}
\label{sec:2HDM}

\subsubsection{The model}

As a final example, we  consider the 2HDM \cite{Lee:1973iz} (for reviews, see e.g. refs.~\cite{Gunion:1989we,Branco:2011iw}); we follow ref.~\cite{Dawson:2022cmu} closely.
The 2HDM is built starting again with the SM (whose Higgs doublet we now identify as $\Phi_1$) and adding a second Higgs doublet, $\Phi_2$.
%
We take their vevs to be $v_1/\sqrt{2}$ and $v_2/\sqrt{2}$, respectively, both of which are assumed to be real. 
A $Z_2$ symmetry is imposed, according to which $\Phi_1 \to \Phi_1, \Phi_2 \to - \Phi_2$. 
If the $Z_2$ symmetry is exact, the model does not have a decoupling limit. The symmetry is assumed to be softly broken, which means that bilinear terms that violate the symmetry are allowed. The potential of the theory thus reads:
\bea
&V_{\textrm{2HDM}}
=\, 
m_{11}^2 \Phi_1^\dagger \Phi_1 + m_{22}^2 \Phi_2^\dagger \Phi_2
- m_{12}^2 \left[\Phi_1^\dagger \Phi_2 + \Phi_2^\dagger \Phi_1 \right]
+ \dfrac{\lambda_1}{2} (\Phi_1^\dagger\Phi_1)^2
+ \dfrac{\lambda_2}{2} (\Phi_2^\dagger\Phi_2)^2
+ \lambda_3 (\Phi_1^\dagger\Phi_1) (\Phi_2^\dagger\Phi_2)&
\no
&+\, \lambda_4 (\Phi_1^\dagger\Phi_2) (\Phi_2^\dagger\Phi_1)
+ 
\dfrac{\lambda_5}{2}
\left[
(\Phi_1^\dagger\Phi_2)^2 + (\Phi_2^\dagger\Phi_1)^2
\right],&
\label{Chap-Real:Vreal}
\eea
with all parameters real.%
\fn{$m_{12}^2$ and $\lambda_5$ can in general be complex. Taking them to be real implies that CP is conserved in the scalar sector of the theory at tree-level. It should be clear, however, that this is but a particular solution of the model with CP violation in the scalar sector, and not a model in itself \cite{Fontes:2021znm}.}
It is convenient to consider a different basis --- the Higgs basis \cite{Donoghue:1978cj,Georgi:1978ri,Botella:1994cs,Branco:1999fs} --- with doublets $H_1$ and $H_2$ defined by:
\be
\left(\begin{array}{c}
H_1 \\
H_2
\end{array}\right)
=
\left(\begin{array}{cc}
c_\beta & s_\beta \\
-s_\beta & c_\beta
\end{array}\right)\left(\begin{array}{c}
\Phi_1 \\
\Phi_2
\end{array}\right),
\ee
with $\beta$ defined such that $t_{\beta}= v_2/v_1$. In the Higgs basis, the potential reads: 
\bea
\label{eq:potential}
V_{\textrm{2HDM}} &=& Y_1 H_{1}^{\dagger} H_{1}
+ Y_2 H_{2}^{\dagger} H_{2} + Y_3 \left( H_{1}^{\dagger} H_{2}+\textrm{h.c.}\right) \no
&&+ \dfrac{Z_{1}}{2}\left(H_{1}^{\dagger} H_{1}\right)^{2}+\dfrac{Z_{2}}{2}\left(H_{2}^{\dagger} H_{2}\right)^{2}+Z_{3}\left(H_{1}^{\dagger} H_{1}\right)\left(H_{2}^{\dagger} H_{2}\right)+Z_{4}\left(H_{1}^{\dagger} H_{2}\right)\left(H_{2}^{\dagger} H_{1}\right) \no
&& + \left\{\dfrac{Z_{5}}{2}\left(H_{1}^{\dagger} H_{2}\right)^{2}+Z_{6}\left(H_{1}^{\dagger} H_{1}\right)\left(H_{1}^{\dagger} H_{2}\right)+Z_{7}\left(H_{2}^{\dagger} H_{2}\right)\left(H_{1}^{\dagger} H_{2}\right)+ \textrm{h.c.}\right\},
\eea
where all parameters are again real. The Higgs doublets are parametrized as:
\be
H_1=\left(\begin{array}{c}
G^{+} \\
\dfrac{1}{\sqrt{2}}\left(v+h_1^{\mathrm{H}}+i G_0\right)
\end{array}\right), \quad H_2=\left(\begin{array}{c}
H^{+} \\
\dfrac{1}{\sqrt{2}}\left(h_2^{\mathrm{H}}+i A\right)
\end{array}\right),
\ee
where $v \equiv \sqrt{v_1^2+v_2^2}=246 \, \mathrm{GeV}$, $H^+$ and $A$ are respectively the charged and CP-odd neutral scalars with masses $m_{H^{\pm}}$ and $m_A$, respectively.
$h_1^{\mathrm{H}}$ and $h_2^{\mathrm{H}}$ do not have well-defined masses; the physical states $h$ and $H$ --- with masses $m=125$ GeV and $m_H$, respectively
--- can be obtained from those two states via the introduction of a new mixing angle, $\alpha$, so that:
\be
\left(\begin{array}{c}
h \\
H
\end{array}\right)=\left(\begin{array}{cc}
s_{\beta-\alpha} & c_{\beta-\alpha} \\
c_{\beta-\alpha} & -s_{\beta-\alpha}
\end{array}\right)\left(\begin{array}{c}
h_1^{\mathrm{H}} \\
h_2^{\mathrm{H}}
\end{array}\right).
\ee
We assume $0 \leq \beta-\alpha \leq \pi$, so that $s_{\beta-\alpha} = \sqrt{1 - c_{\beta-\alpha}^2} > 0$.
In the following, we take all heavy masses to be degenerate and define $M \equiv m_{H} = m_A = m_{H^{\pm}}$.%
\fn{For details on the general case of non-degenerate heavy masses, see ref. \cite{Dawson:2023ebe}.}
One possible choice of independent parameters of the scalar sector is the following:
\be
\label{eq:1-indep-set-2HDM}
c_{\beta \! - \! \alpha},
\,
\beta,
\,
v,
\,
m,
\,
Y_2,
\,
M.
\ee
Another choice  is the set which replaces $Y_2$ by $m_{12}^2$ as an independent parameter: 
\be
\label{eq:2-indep-set-2HDM}
c_{\beta \! - \! \alpha},
\,
\beta,
\,
v,
\,
m,
\,
m_{12}^2,
\,
M.
\ee
The $Z_2$ symmetry is extended to the fermionic sector to avoid flavor-changing neutral currents at tree-level. This extension can be applied in four different ways, each one leading to a different type of 2HDM (for details, see e.g. ref. \cite{Dawson:2022cmu}). In this paper, we consider only the Type-I, as it is the most interesting type from an EFT perspective \cite{Dawson:2022cmu}. Two Feynman rules that are relevant in the discussion of section~\ref{sec:results-2HDM} are:

\vs{5.5mm}
\begin{minipage}[t]{1\linewidth}
	\vs{1.5mm}
	\begin{small}
		\begin{equation}
  \label{eq:2HDM-rules}
			\hspace{14mm}
			-i \dfrac{m_b}{v} \dfrac{1}{\tan \beta} \left(c_{\beta-\alpha} + s_{\beta-\alpha} \, \tan\beta  \right),
            \hspace{45mm}
            i \dfrac{2 m_W^2}{v} s_{\beta-\alpha}  \, g_{\rho\nu}. 
		\end{equation}
	\end{small}
	\begin{picture}(0,30)
	\put(-1,0){\includegraphics[width=0.18\textwidth]{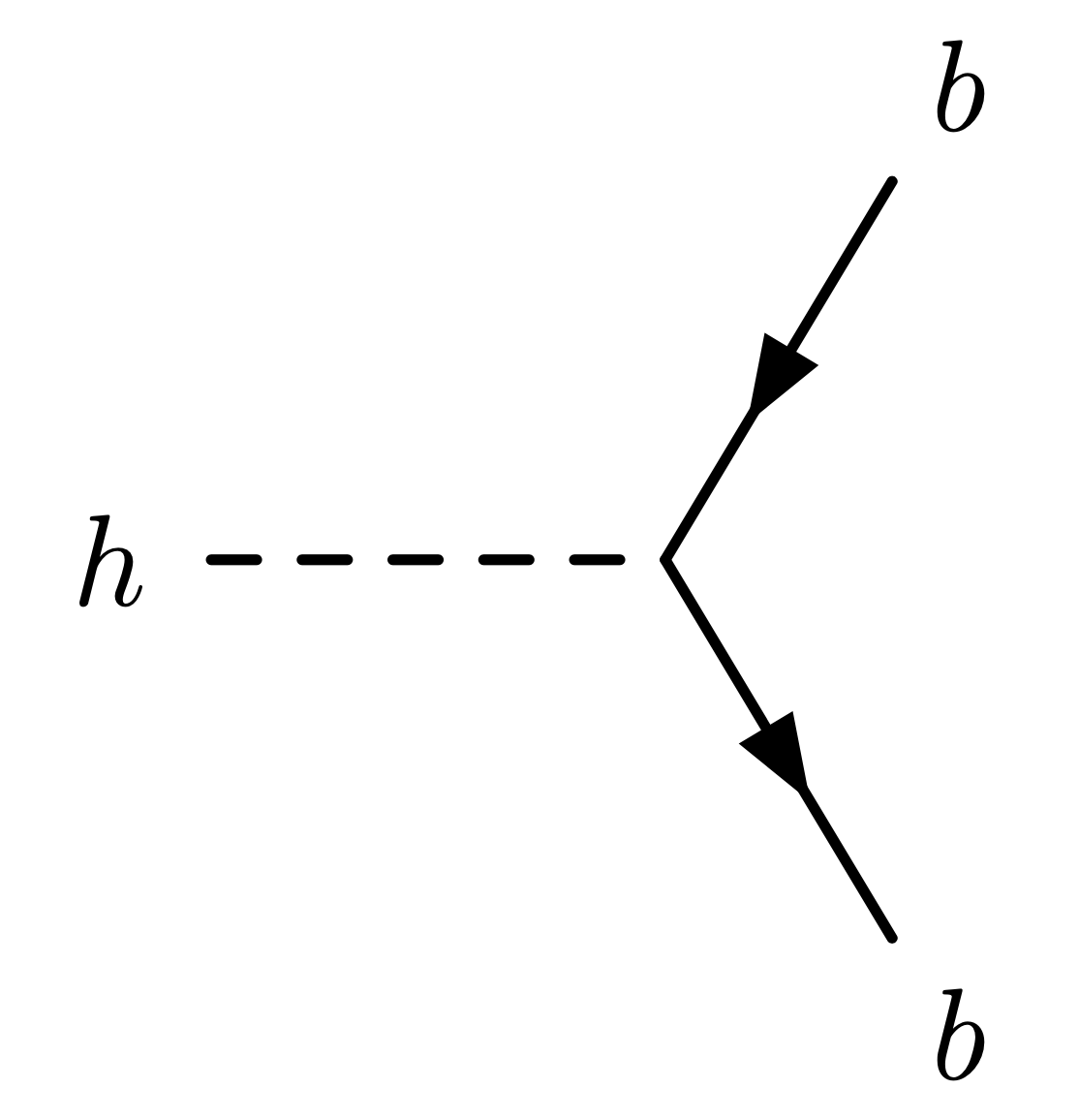}}
	\end{picture}
 \hspace{83mm}
	\begin{picture}(0,30)
	\put(-1,0){\includegraphics[width=0.2\textwidth]{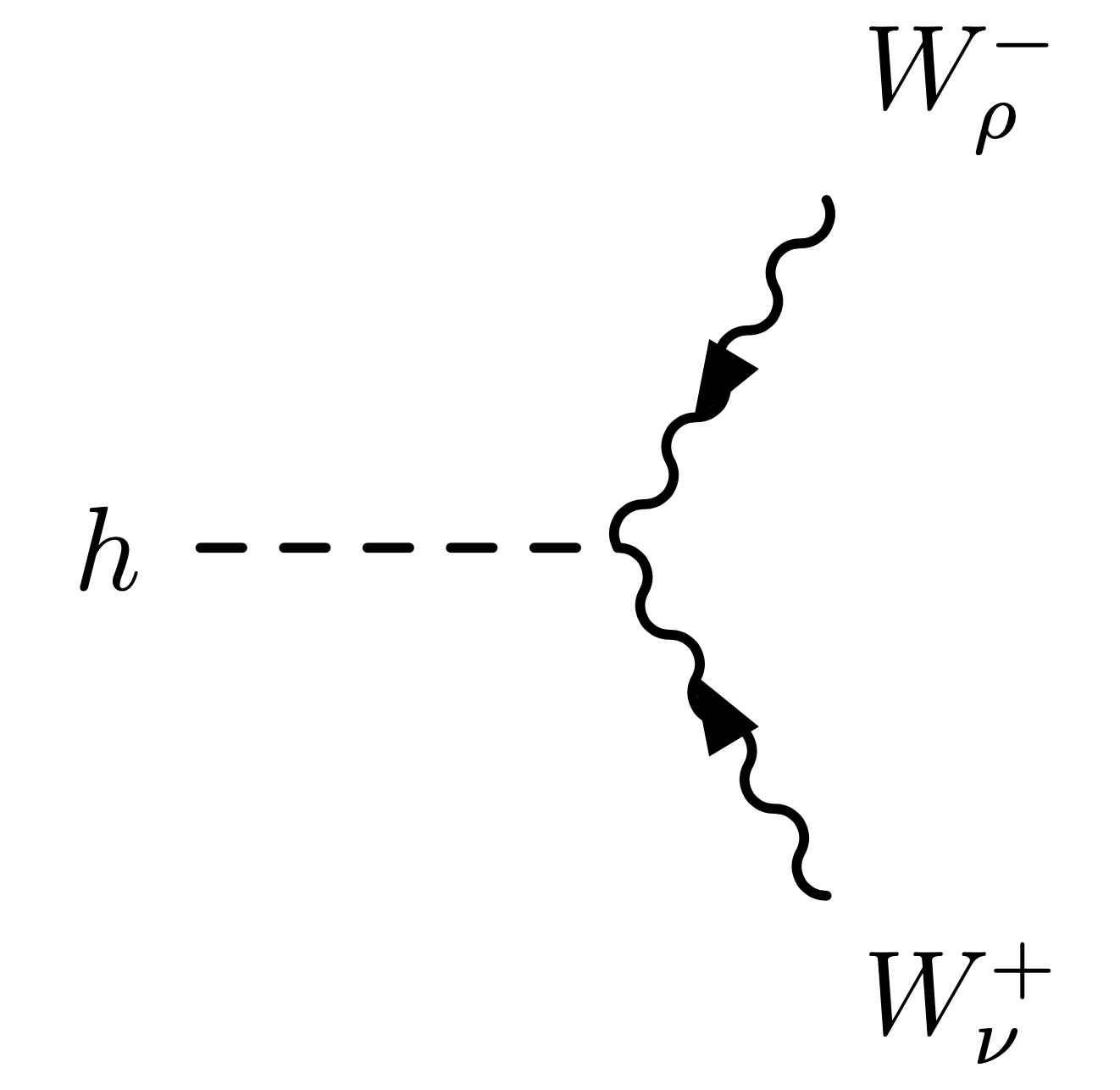}}
	\end{picture} 
\end{minipage}
\vs{-0.5mm}

\noindent Finally, the expressions for $h \to \gamma\gamma$ and $ h \to \gamma Z$ are can be found e.g. in refs. \cite{Fontes:2014xva} and \cite{Gunion:1989we}.

\subsubsection{Restrictions}

The 2HDM is experimentally constrained by many results, including LHC data for the 125 GeV scalar, searches for heavy scalars, Higgs coupling measurements and B meson decays\cite{Anisha:2023vvu}. Constraints resulting from  the contribution of $H^+$ to $b \to s \gamma$ force $\tan \beta \geq  1.2$ \cite{Haller:2018nnx}. 
On the theoretical side, we restrict ourselves to regions of the parameter space complying with boundedness of the potential from below,
perturbative unitarity\cite{Kanemura:1993hm,Akeroyd:2000wc,Ginzburg:2005dt}
and precision electroweak measurements.
We use the analysis of ref. \cite{Dawson:2023ebe} to determine the maximum value of $|c_{\beta-\alpha}|$ allowed for the different values of $\tan \beta$, $Y_2$ and $M$.

\subsubsection{Power countings}

We consider the following possible PCs for an EFT matched to the 2HDM:
\begin{itemize}
\item $\PCT_1$ takes eq.~(\ref{eq:1-indep-set-2HDM}) as the set of independent parameters, and imposes the decoupling scaling:
\be
Y_2 \sim \mathcal{O}(\xi^{-1}),
\qquad
M^2 = Y_2 + \mathcal{O}(\xi^{0}) \sim \mathcal{O}(\xi^{-1}),
\qquad
c_{\beta-\alpha} \sim \mathcal{O}(\xi).
\label{eq:2HDM:PC1}
\ee
\item $\PCT_2$ takes eq.~(\ref{eq:1-indep-set-2HDM}) as the set of independent parameters, and imposes:
\be
Y_2 \sim \mathcal{O}(\xi^{-2}),
\qquad
M^2 \sim \mathcal{O}(\xi^{-2}),
\qquad
c_{\beta-\alpha} \sim \mathcal{O}(\xi).
\label{eq:2HDM:PC2}
\ee
\item $\PCT_3$ takes eq.~(\ref{eq:2-indep-set-2HDM}) as the set of independent parameters, and imposes:
\be
M^2 \sim \mathcal{O}(\xi^{-1}).
\label{eq:2HDM:PC3}
\ee
\end{itemize}
As before, all PCs are theoretically consistent, as they do not induce negative powers of $\xi$ in the effective Lagrangian. $\PCT_1$ is the PC used in ref. \cite{Dawson:2023ebe} and is the decoupling PC\cite{Gunion:2002zf,Carena:2013ooa}; for details, see ref. \cite{Dawson:2023ebe}. Given that $Y_2$ is the parameter that ensures decoupling when taken to be very large, its choice as an independent parameter is reasonable in the decoupling scenario. This choice requires $c_{\beta-\alpha}$ to be small, or else the $h^3$ coupling would scale with negative powers of $\xi$ \cite{Dawson:2023ebe}. This means, in particular, that the choice of eq.~(\ref{eq:1-indep-set-2HDM}) as the set of independent parameters does not allow a consistent expansion solely in terms of $M$ \cite{Dawson:2023ebe}. 
$\PCT_2$ is similar to $\PCT_1$, except that $Y_2$ and the physical heavy masses squared are required to be enhanced by an extra inverse power of the expansion parameter $\xi$.
Finally, $\PCT_3$ is the PC put forward in ref. \cite{Arco:2023sac}; it takes advantage of the set of independent parameters~(\ref{eq:2-indep-set-2HDM}) to avoid scaling $c_{\beta-\alpha}$. With eq.~(\ref{eq:2-indep-set-2HDM}), indeed, the $h^3$ coupling is well-behaved, as it does not depend on positive powers of $M$. This allows an EFT expansion solely in terms of $M$. In particular, $c_{\beta-\alpha}$ does not scale with $\xi$ in $\PCT_3$. 
Note also that, in the HEFT Lagrangian with $\PCT_3$, the $h \bar{b} b$ coupling is identical to that of the full 2HDM model (as a consequence not only of the Feynman rule for that interaction in eq.~(\ref{eq:2HDM-rules}), but also of the fact that $\PCT_3$ does not scale $c_{\beta-\alpha}$).

The matching of the 2HDM to HEFT for the PCs of eqs.~(\ref{eq:2HDM:PC1}), (\ref{eq:2HDM:PC2}) and~(\ref{eq:2HDM:PC3}) is given in tables~\ref{tab:coeffs_2HDM_a} and~\ref{tab:coeffs_2HDM_b}, where we define $\bar{m}_{12}^2 \equiv \mbarh/(m_h^2 s_{\beta} c_{\beta})$ and $t_{\beta} \equiv \tan \beta$. In $\PCT_1$, we write the expressions in the general case where the heavy masses are not necessarily degenerate. More specifically, we follow ref.~\cite{Dawson:2023ebe} and introduce the real quantities $\Delta m_{H}^2$, $\Delta m_{A}^2$ and $\Delta m_{H^{+}}$, such that
$m_{H}^2 = Y_2 + \Delta m_{H}^2, m_{A}^2 = Y_2 + \Delta m_{A}^2, m_{H^{+}}^2 = Y_2 + \Delta m_{H^{+}}^2$.
The results for $\PCT_3$ agree with those of ref. \cite{Arco:2023sac}. 
\begin{landscape}
\renewcommand{\arraystretch}{1.5}
\begin{table}[!h]
\begin{normalsize}
\centering
\begin{tabular}
{>{\centering\arraybackslash}m{0.6cm}
>{\centering\arraybackslash}m{3.8cm}
>{\centering\arraybackslash}m{7.6cm}
>{\centering\arraybackslash}m{9.5cm}
}
\hlinewd{1.1pt}
PC  & $ \Delta b $ & $\Delta \kappa_3$ & $\Delta \kappa_4$ \\
\hline \\[-1.5mm]
$ \PCT_1 $
&
$- \xi^2 \, 3 c_{\beta-\alpha}^2 $ 
&
$-\xi \, 2 c_{\beta-\alpha}^2 \dfrac{Y_{2}}{m_h^2} + \xi^2 \dfrac{1}{2}c_{\beta-\alpha}^2$
&
$- \xi \, 12 c_{\beta-\alpha}^2 \dfrac{Y_{2}}{m_h^2} 
\, +\, \xi^2 c_{\beta-\alpha}^2  \left(\dfrac{16\Delta m_{H}^2}{m_h^2}-11\right)  $ \\ \\\arrayrulecolor{mygray}\hline\\[-1.5mm] 
$\PCT_2$
&
$ -\xi^2 \, 3 c_{\beta-\alpha}^2$
&
$- \dfrac{2 Y_2 c_{\beta -\alpha }^2}{m_h^2} + \xi \dfrac{c_{\beta -\alpha }^3 }{m_h^2 t_{\beta }} (t_{\beta }^2-1 ) (Y_2-M^2 ) \linebreak + \xi^2 \dfrac{ c_{\beta -\alpha }^2 }{2 m_h^2 t_{\beta }^2} \bigg(c_{\beta -\alpha }^2 \Big[M^2 (t_{\beta }^4-4 t_{\beta }^2+1 )+2 Y_2 t_{\beta }^2 \Big] \linebreak + m_h^2 t_{\beta }^2 \bigg) $
&
$\dfrac{4 Y_2 c_{\beta -\alpha }^2}{m_h^2 M^2} \left(M^2-4 Y_2\right) \linebreak + \xi \dfrac{2c_{\beta -\alpha }^3}{m_h^2 M^2 t_{\beta }} \left(t_{\beta }^2-1\right) \left(M^2-12 Y_2\right) \left(M^2-Y_2\right) \linebreak + \mathcal{O}(\xi^2)$ \\ \hline 
$\PCT_3$ &  $ c_{\beta-\alpha}^2 \Big(1 - 2c_{\beta-\alpha}^2 \linebreak + 2 c_{\beta-\alpha}s_{\beta-\alpha}\cot2\beta\Big) \linebreak + \mathcal{O}(\xi)$
&
$-1+ \sba(1+2\cba^2) +\cba^2 \Big[ -2\sba\mbarh \linebreak + 2\cba\cot2\beta \big(1-\mbarh\big) \Big] $
&
$\dfrac{\cba^2}{3}\bigg( -7+64\cba^2-76\cba^4 +12\big(1-6\cba^2+  6\cba^4\big)\bar{m}_{12}^2 \linebreak  + 4\cba\sba\cot 2\beta \Big[-13+38\cba^2-3(-5+12\cba)\bar{m}_{12}^2\Big] 
  \linebreak +4\cba^2\cot^2 2\beta \Big[3\cba^2-16\sba^2+3(-1+6\sba^2)\bar{m}_{12}^2\Big] \bigg) \linebreak + \mathcal{O}(\xi)$ \\
\hlinewd{1.1pt}
\end{tabular}
\end{normalsize}
\caption{$b$, $\kappa_3$ and $\kappa_4$ HEFT couplings for the 2HDM. All the couplings are shown up to $\mathcal{O}(\xi^2)$, except the ones that are too lengthy. The full expressions can be found in the auxiliary file.}
\label{tab:coeffs_2HDM_a}
\end{table}

\vspace{5mm}

\begin{table}[!h]
\begin{normalsize}
\centering
\begin{tabular}
{
> {\centering\arraybackslash}m{1cm}
>{\centering\arraybackslash}m{1.4cm}
>{\centering\arraybackslash}m{4.0cm}
>{\centering\arraybackslash}m{5.8cm}
>{\centering\arraybackslash}m{9.0cm}
}
\hlinewd{1.1pt}
PC
&
$\Delta a$
&
$\Delta c_1$
&
$a_{h \gamma \gamma}$
&
$a_{h \gamma Z}$
\\
\hline\\[-1.5mm]
$\PCT_1$
&
$-\xi^2 \, \dfrac{\cba^2}{2}$
&
$ \xi \dfrac{c_{\beta-\alpha}}{\tan\beta}  - \xi^2 \dfrac{c_{\beta-\alpha}^2}{2}$
&
$-\xi \dfrac{\Delta m^2_H}{48 \pi^2 Y_2}+ \mathcal{O}(\xi^2)$
&
$\xi \dfrac{\Delta m^2_{H^+} (m_Z^2 -2 m_W^2)}{96 \pi^2 m_W^2 Y_2} + \mathcal{O}(\xi^2)$
 \\ \\\arrayrulecolor{mygray}\hline\\[-1.5mm]
$\PCT_2$
&
$-\xi^2 \dfrac{\cba^2}{2}$
&
$ \xi \dfrac{c_{\beta-\alpha}}{\tan\beta} - \xi^2 \dfrac{c_{\beta-\alpha}^2}{2}$
&
$\dfrac{1}{48 \pi ^2 Y_2}  (Y_2-M^2) \linebreak + \xi \dfrac{\cba \cot 2\beta}{48 \pi^2 M^2}(Y_2-M^2) + \mathcal{O}(\xi^2)$
&
$-\dfrac{\left(M^2-\text{Y2}\right) \left(2 m_W^2-m_Z^2\right)}{96 \pi ^2 M^2 m_W^2}-\xi \dfrac{\cot (2 \beta ) \left(M^2-Y_2\right) c_{\beta -\alpha } \left(2 m_W^2-m_Z^2\right)}{96 \pi ^2 M^2 m_W^2}+ \mathcal{O}(\xi^2)$
\\ \\\arrayrulecolor{mygray}\hline\\[-1.5mm]
$\PCT_3$
&
$\sba-1$
&
$ \dfrac{1}{\tan \beta} \Big(c_{\beta-\alpha} \linebreak + \tan\beta \, s_{\beta-\alpha} \Big) -1 $ 
&
$-\dfrac{s_{\beta-\alpha}}{48 \pi^2}+ \xi \dfrac{m_{12}^2}{48 \pi ^2 M^2} \csc (\beta ) \sec (\beta ) \Big(\cot (2 \beta ) c_{\beta -\alpha }+ s_{\beta -\alpha }\Big)-\xi \dfrac{m_h^2}{1440 \pi ^2 M^2} \Big(30 \cot (2 \beta ) c_{\beta -\alpha }+19 s_{\beta -\alpha
   }\Big)+ \mathcal{O}(\xi^2)$ & $\dfrac{s_{\beta-\alpha}}{96 m_W^2 \pi^2} (m_Z^2 - 2 m_W^2) -\xi \dfrac{  \left(2 m_W^2-m_Z^2\right)}{2880 \pi ^2 M^2 m_W^2}\Big[ 30 \cot (2 \beta ) c_{\beta -\alpha } (m_h^2-m_{12}^2 \csc (\beta ) \sec (\beta ))+s_{\beta -\alpha } (19 m_h^2-30 m_{12}^2 \csc (\beta ) \sec (\beta )+2 m_Z^2)\Big]+ \mathcal{O}(\xi^2)$ \\
\\
\hlinewd{1.1pt}
\end{tabular}
\end{normalsize}
\caption{$a$ , $a_{h\gamma \gamma}$ and $a_{h \gamma Z}$ HEFT couplings for the 2HDM. All the couplings are shown up to $\mathcal{O}(\xi^2)$, except the ones that are too lengthy. The full expressions can be found in the auxiliary file.}
\label{tab:coeffs_2HDM_b}
\end{table}
\end{landscape}

\section{Numerical results}
\label{sec:results}

The results that follow were obtained via \textsc{FeynMaster}~\cite{Fontes:2019wqh,Fontes:2021iue} (and its accompanying software~\cite{Christensen:2008py,Alloul:2013bka,Nogueira:1991ex,Mertig:1990an,Shtabovenko:2016sxi,Shtabovenko:2020gxv}), as well as \textsc{FeynArts} \cite{Hahn:2000kx} and \textsc{FormCalc} \cite{Hahn:1998yk}.
The results will be shown in the range allowed by the theoretical constraints of the model being considered.
For each model, we compare the results of the full UV model with those of the PCs introduced in the previous section. Unless mentioned otherwise, all results coincide in the alignment limit ($s_{\chi}=0$ in the Z2RSE, $s_{1}=0$ in the CSE and $c_{\beta-\alpha}=0$ in the 2HDM), which is also the SM result.%
\fn{The only exceptions will be the loop processes $h \to \gamma \gamma$ and $h \to \gamma Z$ in the 2HDM.}

\subsection{Z2RSE}

\begin{figure}[htb!]\vspace{-5mm}
\centering
\includegraphics[width=1\textwidth]{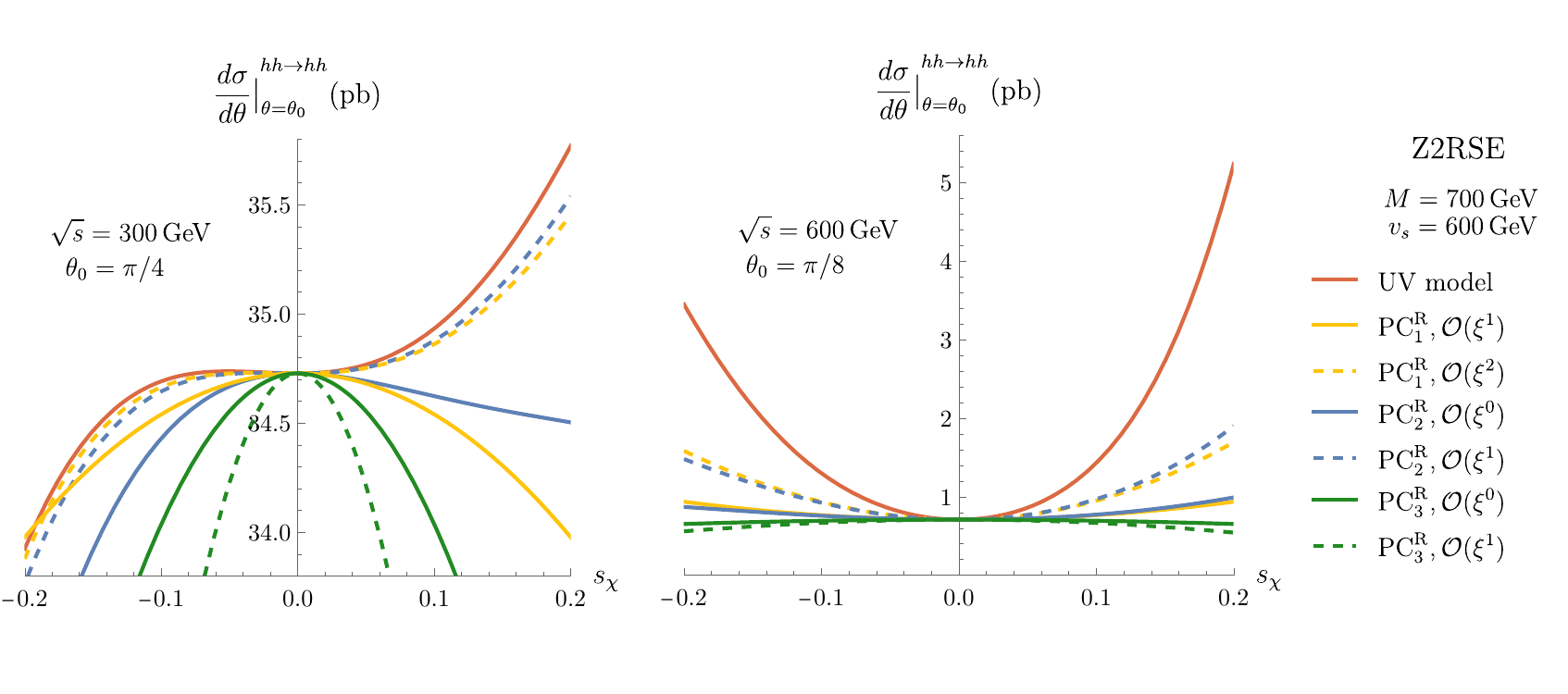}
\vspace{-10mm}
\caption{Comparison between the complete Z2RSE and the HEFT approaches to it in the differential cross-section of $hh \to hh$, for a center-of-mass energy $\sqrt{s}$ and a scattering angle $\theta_0$. 
}
\label{fig:Z2RSE_1}
\end{figure}

In fig.~\ref{fig:Z2RSE_1}, we present the differential cross section $hh \to hh$, for two different values of the center of mass energy: $\sqrt{s} = 300$ GeV (left panel) and $\sqrt{s} = 600$~GeV (right panel).
Let us start by discussing the left panel. We find two very relevant features. The first one concerns the difference between $\PCR_2$ and $\PCR_3$. Recall that these PCs both perform an expansion simply in terms of the heavy mass $M$, and differ only in the choice of independent parameters. Still, the left panel of fig.~\ref{fig:Z2RSE_1} shows that they lead to radically different results. Indeed, whereas $\PCR_2$ clearly converges to the UV model result as higher orders in $\xi$ are considered (such that the $\mathcal{O}(\xi^1)$ result is a very good replication), $\PCR_3$ is utterly unable to properly describe the full model away from the alignment limit, $s_{\chi}=0$, for the both orders shown.%
\fn{We checked that the results eventually improve when $\PCR_3 \, \mathcal{O}(\xi^2)$ is considered.}
This demonstrates not only that the HEFT matching is not unique, but also that making different choices of independent parameters may be crucially important.

Also very interesting on the left panel of fig.~\ref{fig:Z2RSE_1} is the comparison between $\PCR_1$ and $\PCR_2$. As discussed above, the latter is the PC introduced in ref. \cite{Buchalla:2016bse} to describe the HEFT approach to the Z2RSE, whereas the former is the decoupling PC (to recap, $\mathcal{O}(\xi^1)$ and $\mathcal{O}(\xi^2)$ of $\PCR_1$ correspond to SMEFT dimension-6 and dimension-8, respectively). 
The panel shows that, in the region of $s_{\chi} <0$, $\PCR_1$ outperforms $\PCR_2$. This holds for both orders shown: the SMEFT dimension-6 (8) result is closer to the Z2RSE one than the $\mathcal{O}(\xi^0)$ $(\mathcal{O}(\xi^1))$ result of $\PCR_2$. 
One should keep in mind, however, that in terms of the $\xi$ expansion, the first non-trivial order in the $\PCR_1$ expansion is one order higher than the first non-trivial order in the $\PCR_2$ expansion  (since $\mathcal{O}(\xi^0)$ in $\PCR_1$ corresponds to the SM case). Nevertheless, this plot provides an example where a SMEFT approach might be more convenient to reproduce the full model than a (HEFT) approach which only scales the physical heavy masses.

The right panel of fig.~\ref{fig:Z2RSE_1} illustrates a scenario where the EFT expansion starts to break down, as the energies of the problem are very close to those of the UV theory.%
\fn{The scattering angle also changes, for illustrative purposes. The conclusions do not change if $\theta_0 = \pi/4$ is used on the right panel.}
Accordingly, whereas the differences between $\PCR_2$ at $\mathcal{O}(\xi^1)$ and the Z2RSE were around $1\%$ on the left plot, the differences are larger than $50\%$ on the right plot. It is also clear that, for $s_{\chi} <0$, $\PCR_1$ still constitutes a more rapid approach to the full model than $\PCR_2$, which also shows that such a conclusion does not depend on the energy of the problem. Finally, the results for $\PCR_3$ are closer to those of the other PCs than on the left panel.

\begin{figure}[htb!]\vspace{-5mm}
\centering
\includegraphics[width=1\textwidth]{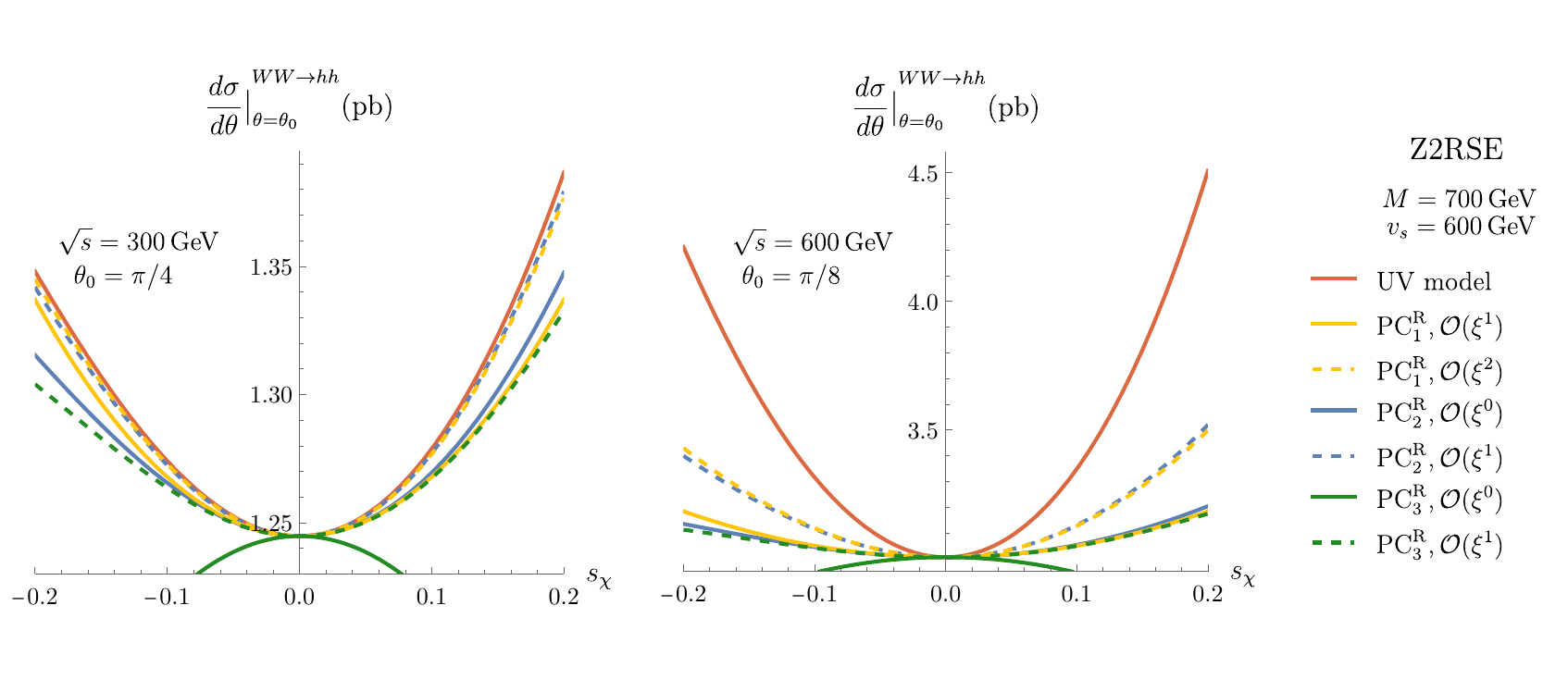}
\vspace{-10mm}
\caption{Comparison between the Z2RSE and HEFT approaches to it in the differential cross-section of $WW \to hh$, for a center-of-mass energy $\sqrt{s}$ and a scattering angle $\theta_0$.}
\label{fig:Z2RSE_2}
\end{figure}

In fig.~\ref{fig:Z2RSE_2}, we present similar plots, but now for the scattering $WW \to hh$. Both panels show essentially the same features as those of the corresponding plot of fig.~\ref{fig:Z2RSE_1} (except that, on the left plot of fig.~\ref{fig:Z2RSE_2}, $\PCR_3$ at $\mathcal{O}(\xi^1)$ now provides a much more faithful description of the full model). This illustrates that the conclusions derived above are not restricted to the process $hh \to hh$. Note also that we investigated different regions of the allowed parameter space, and did not find substantially different conclusions.

\subsection{CSE}

\begin{figure}[htb!]\vspace{-5mm}
\centering
\includegraphics[width=1\textwidth]{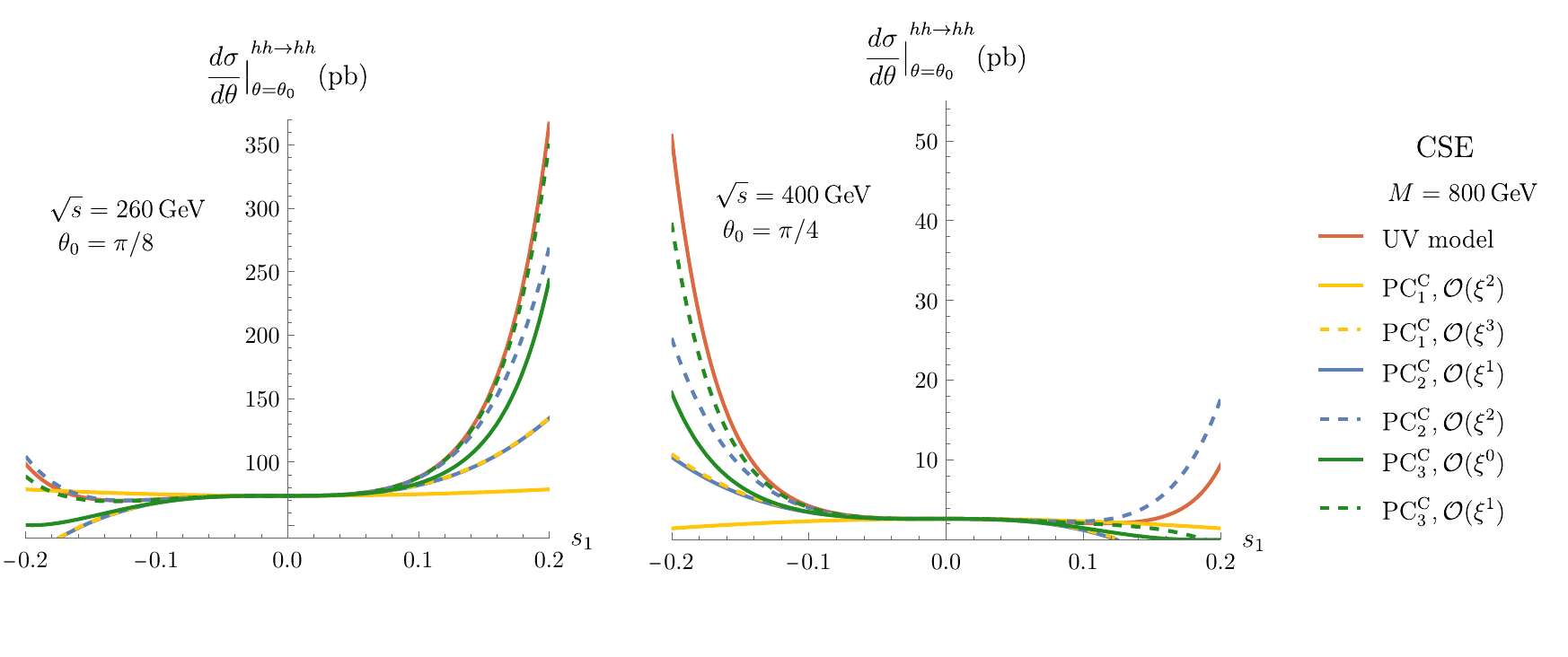}
\vspace{-10mm}
\caption{Comparison between the CSE and HEFT approaches to it in the differential cross-section of $hh \to hh$, for a center-of-mass energy $\sqrt{s}$ and a scattering angle $\theta_0$. On both panels, we use $\alpha_2 = 0, \delta_2 = 0 \, 765, \delta_{3} = 0.695 + i \, 0.145, d_{1} = 0.695 - i \, 7.63, d_2 = 10.6, d_{3} = 1.74 - i \, 4.77, e_{1} = -(28.3 + i \, 20.4) v, e_{2} = -(36.7 - i \, 68.7) v$. This point is derived from a benchmark point provided in ref. \cite{Adhikari:2022yaa}.}
\label{fig:CSE_1}
\end{figure}

We now discuss the same two scattering processes, but in the context of the CSE. In fig.~\ref{fig:CSE_1}, we show $hh \to hh$; the two panels consider different points of the parameter space and different scattering conditions.
Both plots show the same general features. In particular, the results of the full model are essentially constant in a broad region around the alignment limit, $-0.1 < s_1 < 0.1$. All the three PCs thus easily replicate those results. Away from that region, however, $\PCC_1$ (the decoupling PC)
displays a very slow convergence to the full model, with both the $\mathcal{O}(\xi^2)$ and $\mathcal{O}(\xi^3)$ yielding a very poor replication of the full model. $\PCC_2$ shows a
faster convergence, although still with large deviations from the results in shown in red. It is only $\PCC_3$ that is able to provide satisfactory results away from the alignment limit, immediately at $\mathcal{O}(\xi^1)$.

\begin{figure}[htb!]\vspace{-5mm}
\centering
\includegraphics[width=1\textwidth]{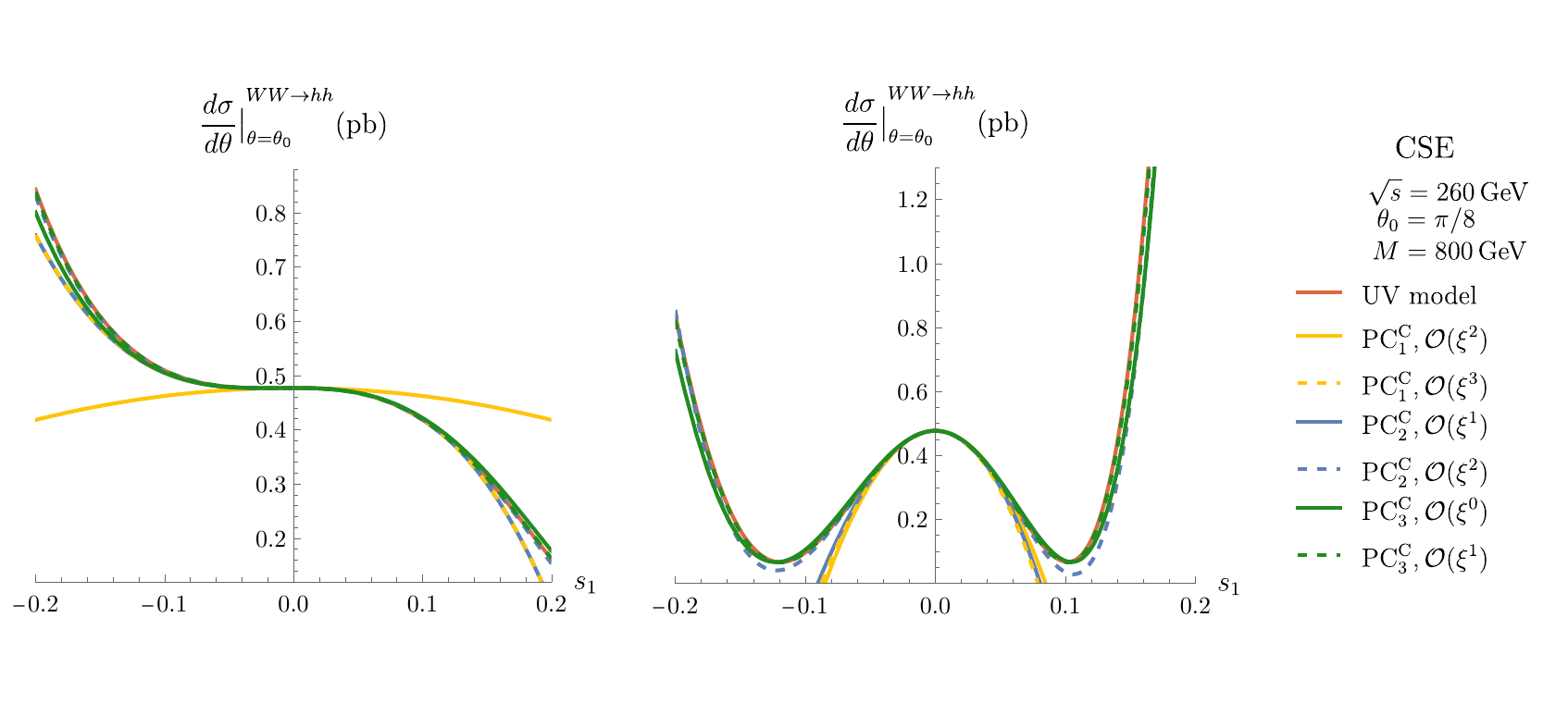}
\vspace{-10mm}
\caption{Comparison between the CSE and HEFT approaches to it in the differential cross-section of $WW \to hh$, for a center-of-mass energy $\sqrt{s}$ and a scattering angle $\theta_0$. The left panel assumes the point of parameter space of fig.~\ref{fig:CSE_1}, whereas the right plot uses $\alpha_2 = 0, d_2 = 0.611, \delta_2 = 24.6, \delta_{3} = 23.5 + i \, 0.00901, d_{1} = -0.0806 + i \, 0.368, d_{3} = -0.128 - i \, 0.0143, e_{1} = -(33 - i \, 28.5) v, e_{2} = - (99.4 + i \, 91.9) v$. As before, this point is derived from a benchmark point provided in ref. \cite{Adhikari:2022yaa}.}
\label{fig:CSE_2}
\end{figure}

Similar conclusions hold also for  $WW \to hh$ scattering, 
which we illustrate in fig.~\ref{fig:CSE_2} for two different points of the parameter space.
Indeed, $\PCC_1$ shows again a very slow convergence away from the alignment limit. For both $hh \to hh$ and $WW \to hh$ scattering, therefore, the SMEFT dimension-8 results (corresponding to $\PCC_1$ at $\mathcal{O}(\xi^2)$) are 
unable to present an accurate description of the full model. Fig.~\ref{fig:CSE_2} demonstrates that $\PCC_2$ has the same pattern in $WW \to hh$ as in $hh \to hh$ scattering, with the  $\mathcal{O}(\xi^2)$ results having small deviations from the full UV results. Finally, $\PCC_3$ is again undisputedly the most adequate PC in both panels of fig.~\ref{fig:CSE_2}, with $\mathcal{O}(\xi^0)$ representing already an excellent description of the CSE results.

\subsection{2HDM}
\label{sec:results-2HDM}

\begin{figure}[htb!]\vspace{3mm}
\centering
\includegraphics[width=0.42\textwidth]{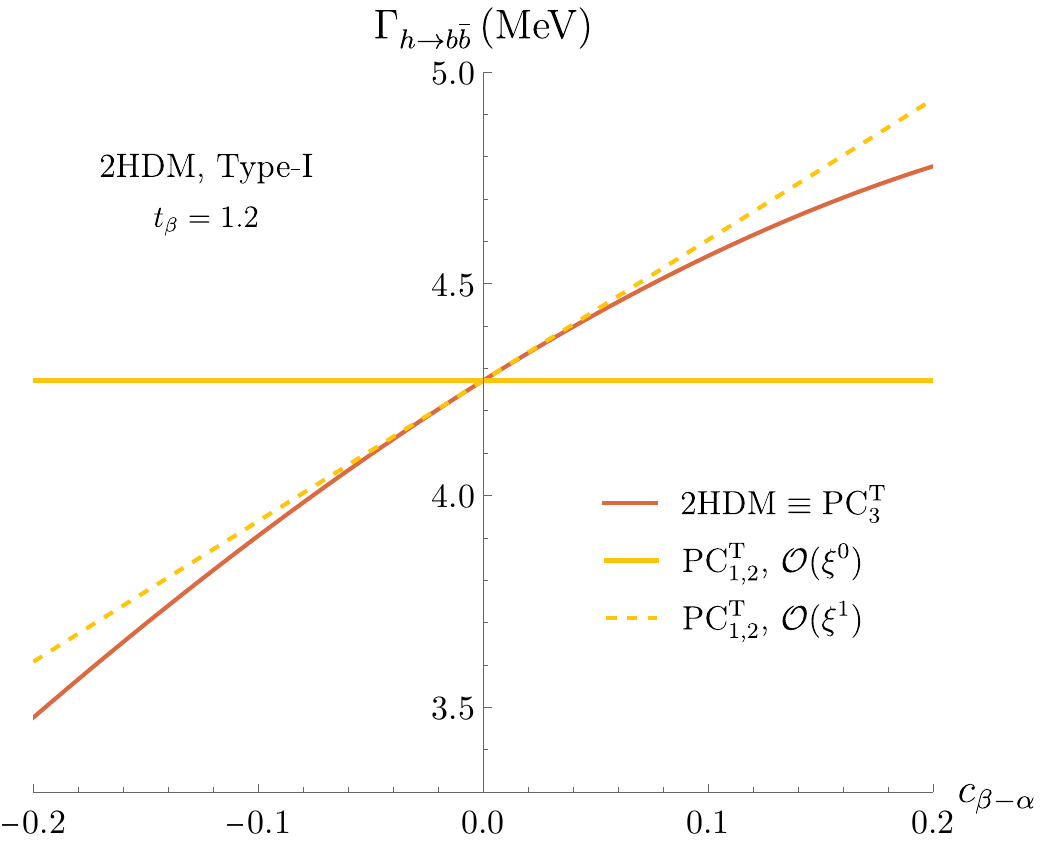}
\hspace{-4mm}
\includegraphics[width=0.58\textwidth]{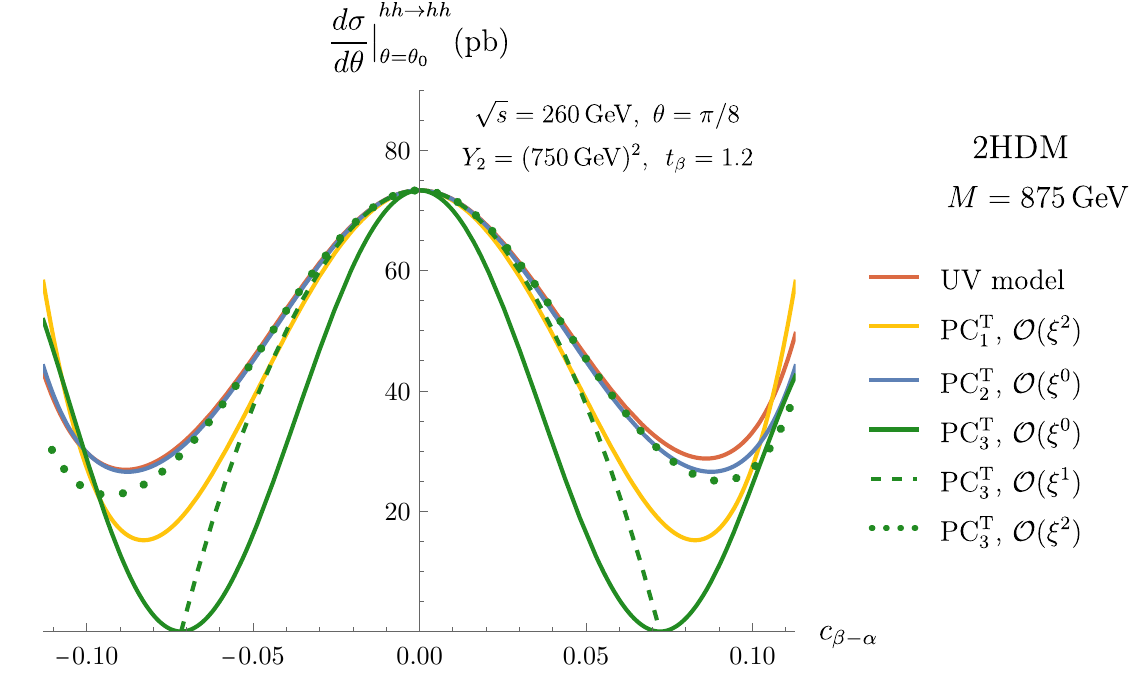}
\vspace{3mm}
\caption{Comparison between the 2HDM and HEFT approaches to it. Left: the decay width of $h \to b \bar{b}$. Right:  differential cross-section of $hh \to hh$.}
\label{fig:2HDM_1}
\end{figure}

In fig.~\ref{fig:2HDM_1}, we show the decay width of $h \to b \bar{b}$ (left panel) and the differential cross-section of $hh \to hh$ (right panel). In both panels, we compare the result in the 2HDM with those of the three PCs introduced in section~\ref{sec:2HDM}. 
%
%
%
In the case of $h \to b \bar{b}$, the $\PCT_3$ result is obviously preferred over the remaining ones; in fact, as discussed in section~\ref{sec:2HDM}, 
it exactly replicates the 2HDM result already at $\mathcal{O}(\xi^0)$.   
$\PCT_1$ and $\PCT_2$ are identical in this process, since only the scaling of $c_{\beta-\alpha}$ determines the result (and that scaling is the same in $\PCT_1$ and $\PCT_2$). For these two PCs, the $\mathcal{O}(\xi^0)$ result is very poor, as it has no dependence on $c_{\beta-\alpha}$, and is thus a constant equal to the 2HDM result for $c_{\beta-\alpha}=0$. In other words, the $\mathcal{O}(\xi^0)$ result of $\PCT_1$ and $\PCT_2$ for $h \to b \bar{b}$ only reproduces the 2HDM in the alignment limit. The $\mathcal{O}(\xi^1)$ result contains a linear dependence on $c_{\beta-\alpha}$, so that it already provides a good reproduction of the 2HDM result within the range of $c_{\beta-\alpha}$ showed.

These results represent a sharp contrast  with those of $hh \to hh$ scattering. In this case, indeed, $\PCT_2$ is by far the most adequate PC. Even though it does not capture the slight asymmetry in $c_{\beta-\alpha}$ of the 2HDM result, it is an excellent approximation to  the latter in the entire range of $c_{\beta-\alpha}$ allowed by the theoretical constraints, and immediatly at $\mathcal{O}(\xi^0)$. By contrast, $\PCT_3$
represents a poor replication of the 2HDM result away from the alignment limit. This holds not only at $\mathcal{O}(\xi^0)$, but also at $\mathcal{O}(\xi^1)$; the latter improves the quality of the replication of the 2HDM result in the range $-0.05 < c_{\beta-\alpha} < 0.05$, but fails to do so in the remaining range. It is only when the $\mathcal{O}(\xi^2)$ truncation is considered that the $\PCT_3$ result properly approaches the 2HDM result away from the alignment limit --- albeit still being worse than $\PCT_2$ at $\mathcal{O}(\xi^0)$.
Finally, the yellow curve represents $\PCT_1$ at $\mathcal{O}(\xi^2)$; this result was first shown in ref. \cite{Dawson:2023ebe}, and shows a relative difference from the 2HDM larger than 40$\%$ for $c_{\beta-\alpha} \simeq 0.08$. The right panel of fig.~\ref{fig:2HDM_1} above thus extends the analysis of that paper, by including two extra PCs.

\begin{figure}[htb!]
\centering
\includegraphics[width=1\textwidth]{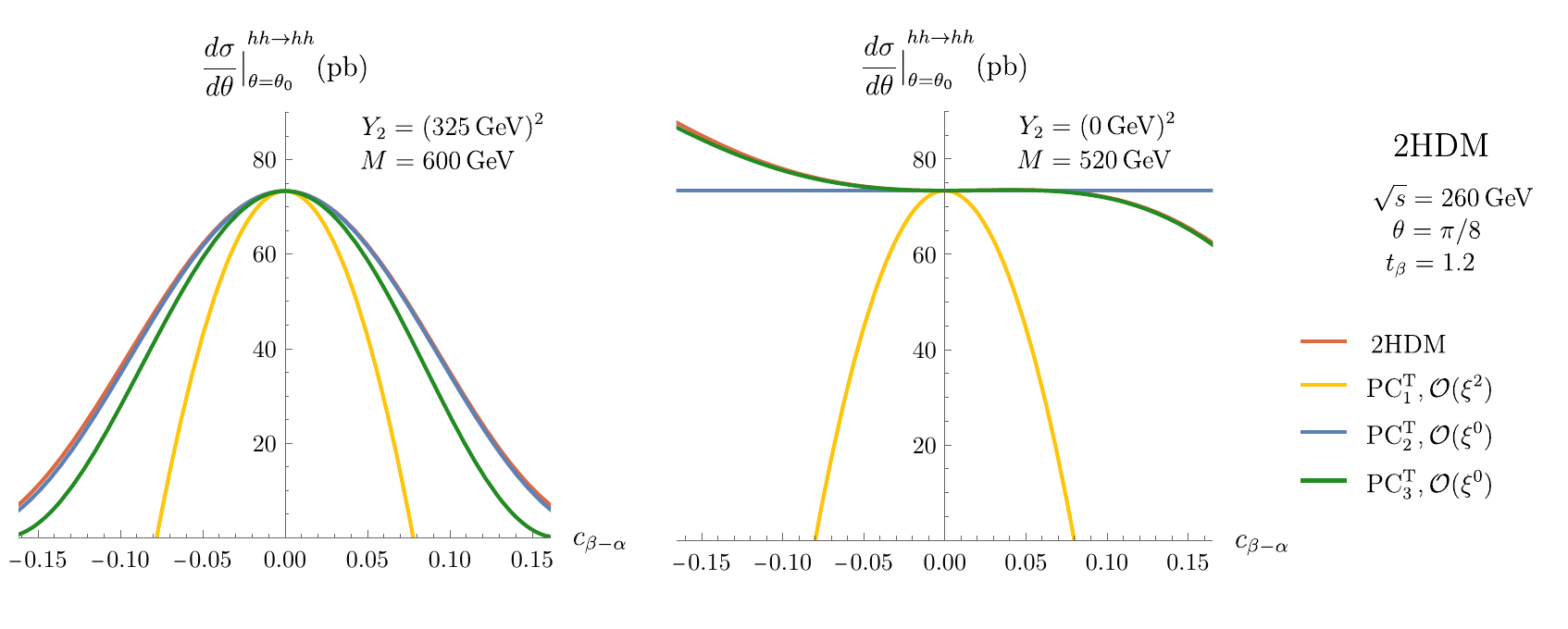}
\vspace{-5mm}
\caption{Comparison between the 2HDM and HEFT approaches to it in the differential cross-section of $hh \to hh$ for non-decoupling scenarios: with $Y_2 = (325 \, \textrm{GeV})^2$ and $M=600 \, \textrm{GeV}$ (left) and $Y_2 = 0$ and $M=520 \, \textrm{GeV}$ (right).}
\label{fig:2HDM_hhhh_more}
\end{figure}

One might wonder if the results of the right panel of fig.~\ref{fig:2HDM_1} depend significantly on the region of parameter space considered. To address this question, we show in fig.~\ref{fig:2HDM_hhhh_more} the same observable, but for very different values of $Y_2$: $(325 \, \textrm{GeV})^2$ on the left panel, and $0$ on right panel.
Both scenarios constitute clear deviations from the decoupling limit $Y_2 \gg v^2$. Therefore, the observed failure on both panels of $\PCT_1$ (the decoupling PC) to reproduce the 2HDM result away from $c_{\beta-\alpha} = 0$ was expected.
As for the other PCs, the figure shows that they are very good replications of the 2HDM. That $\PCT_2$ has such accurate result is surprising, as it imposes the scaling $Y_2 \sim \mathcal{O}(\xi^{-2})$, which thus seems incompatible with very small values of $Y_2$. On the other hand, the circumstance that this PC simultaneously imposes the scalings $c_{\beta-\alpha} \sim \mathcal{O}(\xi)$ and $M^2 \sim \mathcal{O}(\xi^{-2})$ compensates for that apparent incompatibility, so that the end result is very accurate. Still, $\PCT_2$ has a less accurate description of the full model than $\PCT_3$ on the right panel of fig.~\ref{fig:2HDM_hhhh_more}. This shows that the conclusions of the right panel of fig.~\ref{fig:2HDM_1} do depend on the region of parameter space considered.

\begin{figure}[htb!]\vspace{-5mm}
\centering
\includegraphics[width=1\textwidth]{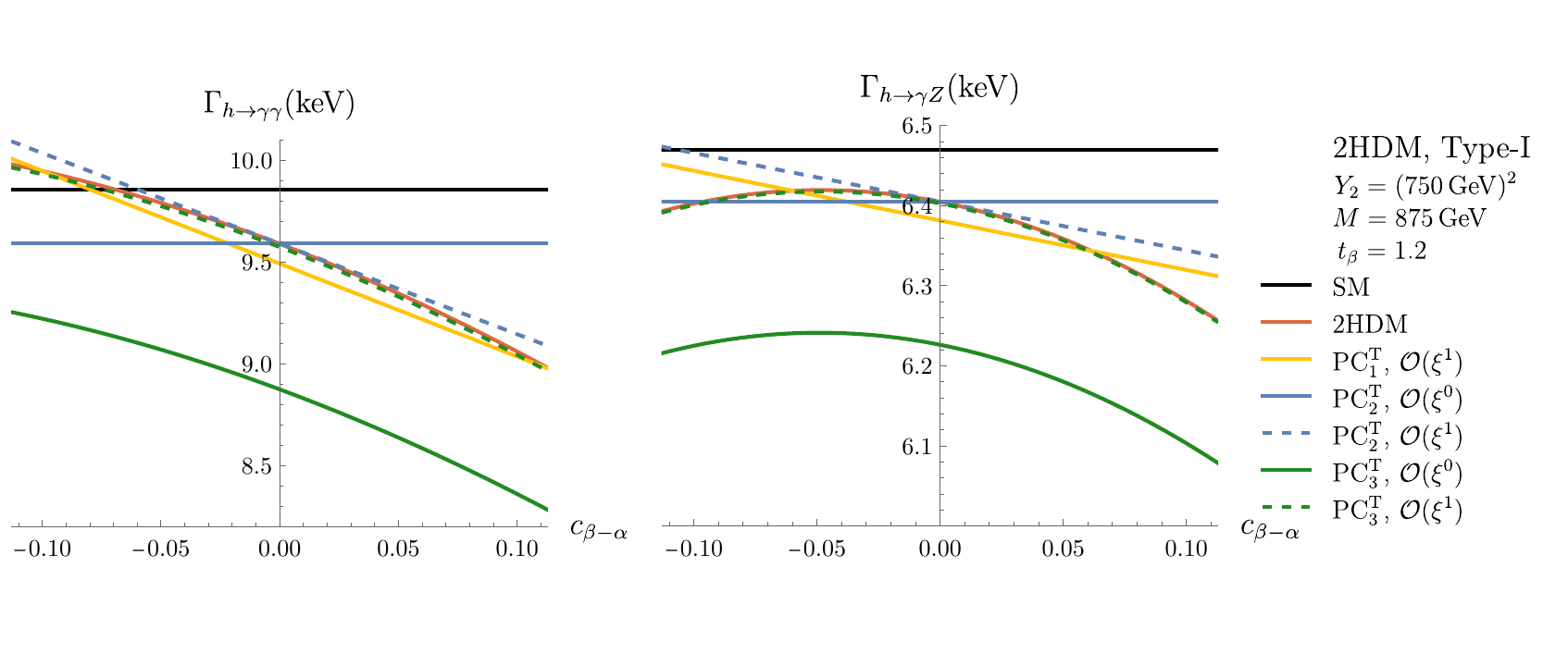}
\vspace{-10mm}
\caption{Comparison between the 2HDM and HEFT approaches to it in the decay width of $h \to \gamma \gamma$ (left) and $h \to \gamma Z$ (right).}
\label{fig:2HDM_hAA_hAZ}
\end{figure}

Fig.~\ref{fig:2HDM_hAA_hAZ} shows the decay width of $h \to \gamma \gamma$ (left panel) and $h \to \gamma Z$ (right panel), again comparing the different possible PCs with the 2HDM result. In both cases, we also show the SM result, which is not obtained by the 2HDM one in the alignment limit $c_{\beta-\alpha}=0$. 
This feature is related to the well-known non-decoupling effects typical of these processes, which are caused by the loop of charged scalars \cite{Djouadi:1996yq,Haber:2000kq,Arhrib:2003vip,Bhattacharyya:2013rya,Ferreira:2014naa,Fontes:2014tga}.
We checked that, in both processes, $\PCT_1$ at $\mathcal{O}(\xi^0)$ generates the SM curve, and not the 2HDM one. This result is in agreement with the circumstance that $\PCT_1$ is the decoupling PC. The figure shows that $\PCT_1$ presents a reasonable description of the 2HDM already at $\mathcal{O}(\xi^1)$. This means that, even though the loops with charged SM bosons do not contribute at this order (recall eq. \ref{eq:2HDM-rules}), the modification of the Yukawa interactions is sufficient to provide a decent replication of the UV model result, albeit with a simple linear dependence on $c_{\beta-\alpha}$. Fig. \ref{fig:2HDM_hAA_hAZ} also shows that $\PCT_3$ at $\mathcal{O}(\xi^0)$ on both panels is quite deviated from the 2HDM result, even in the alignment limit. That curve has an approximately constant relative difference with regards to the 2HDM throughout all the allowed range of $c_{\beta-\alpha}$: around $7\%$ in $h \to \gamma \gamma$ and around $3\%$ in $h \to \gamma Z$. On the other hand, $\PCT_3$ at $\mathcal{O}(\xi^1)$ describes an excellent replication of the 2HDM result in both processes. This is not what happens in $\PCT_2$, which, despite presenting a smaller relative difference to the 2HDM result at $\mathcal{O}(\xi^0)$ (in its constant value), does not provide as adequate an approach to the 2HDM result as $\PCT_3$ at $\mathcal{O}(\xi^1)$.

\section{Conclusions}
\label{sec:conclusions}

The HEFT can be used at the LHC to parametrize possible deviations from the SM. Eventual non-zero HEFT coefficients should then be converted into coefficients of specific UV models via a matching procedure. In this paper, we noted that such a procedure is not unambiguous. Even if a simple expansion in inverse powers of a heavy mass is used, very different power countings (PCs) are obtained by considering different set of independent parameters. Moreover, different PCs approach the full model differently for different observables.
We illustrated these aspects by considering three BSM models with an extended scalar sector: the $Z_2$-symmetric real singlet extension (Z2RSE), the complex singlet extension (CSE) and the 2-Higgs Doublet Model (2HDM). For each model, we investigated three PCs, providing the relevant HEFT coefficients. In all the models, one of the PCs chosen is the decoupling PC, in the sense that it leads to the same results as the SMEFT.

In the Z2RSE, we showed that two PCs that perform an expansion just in inverse powers of the heavy physical mass lead to very different results, due to different choices for the independent parameters. We also showed regions where both approach the full theory less rapidly than the SMEFT results. In the CSE, the decoupling PC leads to a very slow convergence, so that even the SMEFT dimension-8 results are inadequate to replicate the full model away from the alignment limit. Finally, in the 2HDM, we illustrated how different PCs are more accurate for different observables. For example, while the PC identified as $\PCT_3$ is most adequate for 2 body decays of the Higgs boson, $\PCT_2$ is clearly more accurate for some regions of processes such as $hh \to hh$. We also showed that an intuition for the performance of a PC is not always available, as $\PCT_2$ properly replicates the 2HDM for very low values of $Y_2$, even though it scales this parameter as a heavy one. As a global conclusion, we note that the existence of multiple theoretically consistent PCs complicates the interpretation of HEFT coefficients in terms of parameters of specific UV models.

This paper is the first exploration of the existence of multiple possible matchings between the HEFT and a UV model. In the future, other models and other processes could be considered, and a systematic comparison between PCs can be put forward. 
It would be also relevant  
to investigate the effect of radiative corrections, and their impact on the consistency of different PCs.
An auxiliary file accompanies this manuscript, containing the HEFT couplings for the CSE and the 2HDM.

\section*{Ackowledgments}

D.F. is grateful to Ilaria Brivio, Supratim Das Bakshi, Howard Haber, Aneesh Manohar and Matthew Sullivan for discussions, as well as to the Mainz Institute for Theoretical Physics (MITP) of the Cluster of Excellence PRISMA+ (Project ID 39083149) for its hospitality and support. C.Q.C and J.J.S.C. are grateful to Gerhard Buchalla for discussions about the singlet extension, and to Francisco Arco, Mar\'\i a J. Herrero and Roberto A. Morales for comments on the 2HDM.
The authors also thank early conversations with Lucas Barbero.
S. D. and D. F. are supported by the U.S. Department of Energy under Grant Contract No. DE-SC0012704. C.Q.C has been funded by the MINECO (Spain) predoctoral grant BES-2017-082408.   
This work was supported in part by the EU under grant 824093 (STRONG2020), Spanish MICINN under PID2019-108655GB-I00/AEI/10.13039/501100011033, PID2019-106080GB-C21, Universidad Complutense de Madrid under research group 910309 and the IPARCOS institute. This preprint has been issued with number IPARCOS-UCM-23-129.

\appendix
\renewcommand{\thesection}{\Alph{section}}
\renewcommand{\theequation}{\thesection.\arabic{equation}}
\setcounter{equation}{0}

\section{HEFT approach to the CSE}

In this appendix, we focus on the integration out of the heavy scalars of the CSE, as has already been done for the Z2RSE \cite{Buchalla:2016bse} and the 2HDM \cite{Dawson:2023ebe,Arco:2023sac}.
We begin by considering the potential in eq.~(\ref{eq:CSE:potential}) in terms of the physical states $h_1$, $h_2$ and $h_3$. We can split the Lagrangian in the form,
\begin{equation}
\mL_{\rm CSE}\,=\, \mL_{\rm CSE}^{\rm light}+\mL_{\rm CSE}^{\rm heavy}\, ,
\end{equation}
where $\mL_{\rm CSE}^{\rm light}$ involves only light (i.e. non-BSM) fields. On the other hand,  
$\mL_{\rm CSE}^{\rm heavy}$ involves light and heavy fields, and we can conveniently write it as follows:
\begin{equation}
\mL_{\rm CSE}^{\rm heavy} = 
\frac{1}{2}(\partial_\mu H^a)^2  -\frac{1}{2} (M^2)^{ab} H^a H^b   
 + J_1^a H^a +J_2^{ab} H^a H^b +J_3^{abc} H^a H^b H^c + J_4^{abcd} H^aH^b H^c H^d\, ,
\label{eq:LCSEjs}
\end{equation}
where $(M^2)^{ab}$ is the diagonal squared mass matrix, $H^a=(h_2 , \, h_3)$ are the heavy fields, and the $J_k$ contain only light fields. 
The heavy scalars $H^a$ are integrated out at tree-level by solving their equations of motion (EoM):
\begin{eqnarray}
\label{eq:eomgeneral}
J_1^a\, +\, (-\partial^2-M^2+2 J_2)^{ab} H^b + 3 J_3^{abc} H^b H^c +4 J_4^{abcd} H^b H^c H^d \,=\, 0\, .
\end{eqnarray}

As we are only interested in at most $2 \rightarrow 2$ scattering processes at tree level,%
\footnote{Notice that for one-loop amplitudes with three and four external particles, vertices with more than four fields are also in general relevant.}%
we neglect terms with more than four fields in $\mL_{\rm CSE}^{\rm light}$. We also neglect EoM solutions that give rise to terms with more than four light fields in the EFT Lagrangian. Since $J_{1}$ and $J_2$ have at least two and one light fields, respectively, the EoM solutions for $H^a=(h_2 , \, h_3)$ with two fields are fully contained in:
\begin{equation}
\overline{H}^a \, =\, \sum_{n=0}^\infty (M^{-2-2n})^{ab} \partial^{2n} J_1^b\, =\, 
(M^{-2})^{ab} J_1^b 
\, + \, 
(M^{-4})^{ab} \partial^2 J_1 \, + \, ... \, ,
\end{equation}
where $J_1$ has terms with two and three light fields. 
Note that the full $H^a$ solution for~(\ref{eq:eomgeneral}) has no terms with zero and one fields (it starts with the two-light-field terms in $\overline{H}^a$). 
One can thus easily observe that the $J_2$ and $J_3$ terms in the Lagrangian~(\ref{eq:LCSEjs}) will give EFT operators with at least five and six fields, as $J_2$, and $J_3$ start with one and zero light fields, respectively. Furthermore, the EFT operators produced by the $J_4$ term in Lagrangian~(\ref{eq:LCSEjs}) will contain at least eight light fields, where $J_4$ starts with zero fields. 
Hence, the resulting EFT operators 
containing up to four light fields are simply provided by the first line of eq.~(\ref{eq:LCSEjs}):%
\fn{This result is actually general. By construction, indeed, the terms of~(\ref{eq:LCSEjs}) that only contain two heavy fields are included in $(M^2)^{ab}$, while the operators that contain at least one light field in addition to the two heavy ones are given in $J_2^{ab}$. 
Moreover, if the $H^a$ particles are mass eigenstates and there are no heavy--light mass mixing terms, $J_1^a$ will always contain at least two light fields. These two are the assumptions that we have employed to extract the 4-field EFT operators in this appendix.}
\begin{equation}
 \mathcal{L}_{\rm CSE}^{\rm EFT} \Big|_{\rm \leq 4\, fields}\, \subset\,  \mathcal{L}_{\rm CSE}^{\rm light}
+\frac{1}{2}(\partial_\mu \overline{H}^a)^2 
 -\frac{1}{2} (M^2)^{ab} \overline{H}^a \overline{H}^b     + J_1^a \overline{H}^a \, .
\label{eq:CSE-HEFT}
\end{equation}
Now, it is possible to match this effective Lagrangian with the one given by HEFT in eq.~(\ref{eq:heftdef}). This matching yields the coefficients $a$, $b$, $\kappa_3$, and $\kappa_4$ of the lowest order HEFT Lagrangian shown in table~\ref{tab:coeffs_CSE}. The expressions for $J_1^a$ up to two light fields are: 
\begin{align}
    J_1^{h_2} = & \dfrac{c_2 s_1}{v} (2 m_W^2 W_{\mu } W_{\mu }^{\dagger} + m_Z^2 Z_{\mu}{}^2) + \dfrac{h_1^2}{48 v} \Big(3 c_2 s_1^3 (-3 v^2 \bar{\delta}_{23} + 4 m_h^2 + 2 m_2^2)  \nonumber \\
& + 3 s_1 v^2 ((9 c_1^2 - 1) c_2 \bar{\delta}_{23} - 8 c_1 \delta_{\mathrm{3I}} s_2) + \sqrt{2} s_1^2 v (2 s_2 (3 e_{\mathrm{1I}} + e_{\mathrm{2I}}) - 9 c_1 c_2 e_{\mathrm{12R}}) \nonumber  \\
& + \sqrt{2} (c_1^2 - 1) v (3 c_1 c_2 e_{\mathrm{12R}} - 2 s_2 (3 e_{\mathrm{1I}} + e_{\mathrm{2I}})) - 6 (3 c_1^2 + 1) c_2 s_1 (2 m_h^2 + m_2^2)\Big),\label{eq:J1h2} \\
J_1^{h_3} = & \dfrac{s_1 s_2}{v} (2 m_W^2 W_{\mu } W_{\mu }^{\dagger} + m_Z^2 Z_{\mu}{}^2)  + \dfrac{h_1^2}{48 v} \Big(-3 s_1 v^2 (-9 c_1^2 s_2 \bar{\delta }_{23} + (3 s_1^2 + 1) s_2 \bar{\delta}_{23} - 8 c_2 c_1 \delta_{\mathrm{3I}}) \nonumber  \\
& + \sqrt{2} v (2 c_2 (c_1^2 - s_1^2 - 1 ) (3 e_{\mathrm{1I}} + e_{\mathrm{2I}} ) + 3 c_1 s_2 (c_1^2 - 3 s_1^2 - 1 ) e_{\mathrm{12 r}} ) \nonumber  \\
& + 6 s_1 s_2 (-3 c_1^2 + s_1^2 - 1 ) (2 m_h^2 + m_3^2 )\Big). \label{eq:J1h3}
\end{align}

\addcontentsline{toc}{section}{References}

\bibliographystyle{h-physrev4}
\bibliography{MyReferences}

\printindex
\end{document}